\documentclass[preprint,11pt]{aastex}

\newcommand{\lam}{$\lambda$}

\newcommand{\ecss}{erg cm$^{-2}$ s$^{-1}$ sr$^{-1}$} 
\newcommand{\ecs}{erg\,cm$^{-2}$\,s$^{-1}$ sr$^{-1}$} 
\newcommand{\kms}{km~s$^{-1}$}

\renewcommand{\ion}[2]{#1\,{\sc #2}}

\newcommand{\hinode}{\emph{Hinode}}

\def\ionx[#1 #2]{#1\,{\sc #2}}

\begin{document}


\title{Velocity measurements for a solar active region fan loop from
  Hinode/EIS observations}


\author{P. R. Young\altaffilmark{1},
B. O'Dwyer\altaffilmark{2}
\and H. E. Mason\altaffilmark{2}
}

\altaffiltext{1}{George Mason University, 4400 University Drive, Fairfax, VA 22030}
\altaffiltext{2}{Department of Applied Mathematics and Theoretical
  Physics, University of Cambridge, Wilberforce Road, Cambridge, CB3 0WA, UK}


\begin{abstract}
The velocity pattern of a fan loop structure within a solar  active region
over the temperature range 0.15--1.5~MK is derived using data from the
EUV Imaging Spectrometer (EIS) on board the \hinode\ satellite. The loop is
aligned towards the observer's line-of-sight and shows
downflows (redshifts) of around 15~\kms\ up to a temperature of 0.8~MK, but for
temperatures of 1.0~MK and above the measured velocity shifts are
consistent with no net flow. This velocity result applies over a
projected spatial distance of 9~Mm and demonstrates that the cooler,
redshifted plasma is physically disconnected from the hotter,
stationary plasma. A scenario in which the fan loops consist of at
least two groups of ``strands'' -- one cooler and downflowing, the other
hotter and stationary -- is suggested. The cooler strands may represent
a later evolutionary stage of the hotter strands. A density diagnostic of
\ion{Mg}{vii} was used to show that the electron density at around
0.8~MK falls from $3.2\times 10^9$~cm$^{-3}$ at the loop base, to
$5.0\times 10^8$~cm$^{-3}$ at a projected height of 15~Mm. A filling
factor of 0.2 is found at temperatures close to the formation
temperature of \ion{Mg}{vii} (0.8~MK), confirming that the cooler,
downflowing plasma occupies only a fraction of the apparent loop
volume. The fan loop is rooted within a so-called ``outflow region''
that displays low intensity and blueshifts of up to 25~\kms\ in 
\ion{Fe}{xii} \lam195.12 (formed at 1.5~MK), in contrast to the loop's
redshifts of 15~\kms\ at 0.8~MK.
A new technique for obtaining an absolute wavelength calibration for
the EIS instrument is presented and an instrumental effect, possibly
related to a distorted point spread function, that affects velocity
measurements is identified.
\end{abstract}


\section{Introduction}

Amongst the complex and varied range of plasma structures in the solar
corona, active region fan loops stand out as large,
long-lived structures.  The loops are most distinctive when observed in emission
lines formed in the narrow temperature range 0.6--1.0~MK, and
they came to prominence when the Transition Region and Coronal
Explorer (TRACE) satellite began regular observations in an extreme
ultraviolet filter centered at 174~\AA\ \citep{handy99}, picking up strong emission
lines of \ion{Fe}{ix} and \ion{Fe}{x} that are formed at 0.8 and
1.0~MK, respectively. This filter is
  referred to as ``TRACE 171'' to be consistent with the terminology
  used for the earlier EUV Imaging Telescope \citep[EIT;][]{delab95}
  instrument on board the Solar and
Heliospheric Observatory \citep[SOHO;][]{domingo95} although  in fact the filter is
  more sensitive
  to \ion{Fe}{x} \lam174.53 \citep{handy99}.
\citet{schrijver99} presented
properties of the active region fan loops seen in the TRACE 171 filter which we
summarize here. They are seen to terminate in the penumbra of sunspots and also in
strong fields at the edges of active regions; they show small 
temperature variation with height; they can survive for hours to days;
and they show propagating intensity perturbations (PIPs) that give the
impression of upflows from the surface. The latter property has been
extensively studied, and the most common interpretation is that PIPs
are due to upward-propagating, slow magnetoacoustic waves
\citep[e.g.,][]{demoortel02,wang09}. Note that the PIPs are not a
permanent feature of fan loops but are intermittent, lasting for
typically 10's of minutes before switching off \citep{mcewan06}. This
work also demonstrated that individual `strands' within the loop can
display PIPs independently.

Prior to TRACE, the presence of spiky, plumelike structures at the
edges of active regions had been remarked on in analyzes of
\emph{Skylab} data. For example, \citet{foukal76} identified \ion{Ne}{vii} ($T=0.6$~MK)
plumes close to a sunspot, while \citet{cheng80} found emission
``spikes'' in \ion{Ne}{vii} at the periphery of an active region, with
\ion{Mg}{ix} ($T=1.0$~MK) loops corresponding with \ion{Ne}{vii} although the
emission was more extended and less sharply defined. These structures
are almost certainly the same as the TRACE fan loops.

The first line-of-sight (LOS) velocity measurements for a fan loop were made by
\citet{winebarger02} using the SUMER \citep[Solar Ultraviolet
Measurements of Emitted Radiation;][]{wilhelm95} instrument on board
SOHO. A TRACE 171 fan loop was clearly 
identified in the \ion{Ne}{viii} \lam770 emission line (0.7~MK) observed by
SUMER and redshifts of 15--40~\kms\ were measured, i.e., plasma is
downflowing in the fan loop legs. This is in striking contrast to the
apparent upflows in the TRACE image sequences first described by
\citet{schrijver99}, and appears to confirm the standard interpretation of the
PIPs as due to magnetoacoustic waves rather than an upward flow of
plasma.  This view has been challenged more recently by 
\citet{depontieu10} who interpret the PIPs as quasi-periodic bursts of
upflowing plasma with velocities of 50--150~\kms. The emission from
this plasma is weak compared to that from bulk loop plasma, but is
sufficient to give an intensity oscillation signal (which can be
detected with an imaging instrument), coupled with
oscillations in velocity and line width.

Further active region velocity studies with SUMER were performed by \citet{marsch04}
and \citet{doschek06}. The former presented Doppler maps for three
active regions \citep[including that of][]{winebarger02} and, although
not specifically remarked on, it is clear that several fan loop
structures (including some close to a sunspot) in the data display
significant redshifts in \ion{Ne}{viii} \lam770. The loops also appear
to show redshifts in \ion{C}{iv} \lam1548. \citet{doschek06} presented
Doppler maps in \ion{C}{iv} \lam1548, \ion{S}{v} \lam786 and
\ion{Ne}{viii} \lam770 (formed at 0.1, 0.2 and 0.7~MK, respectively)
for an active region and two ``plumelike'' 
structures (almost certainly the same as fan loops) were shown to
exhibit redshifts of 5--10~\kms\ in \ion{S}{v} and \ion{Ne}{viii}
although in this case not in \ion{C}{iv}. The SUMER instrument had
limited access to coronal emission lines, and no velocity studies of
fan loops were made for higher temperatures than \ion{Ne}{viii} (0.7~MK).

The EUV Imaging Spectrometer \citep[EIS;][]{culhane07} on board the
\hinode\ satellite \citep{kosugi07}  is the first solar ultraviolet spectrometer
to routinely allow the measurement of coronal emission line Doppler
shifts to precisions of 0.5~\kms\ \citep{mariska08}. It also has access to lines from the
upper transition region (temperatures 0.2--1.0~MK) and thus there is
some temperature overlap with the SOHO/SUMER instrument. EIS
therefore, for the first time, allows the change in velocity structure
from the transition region to 
the corona to be investigated with a single instrument. Studies of
\citet{delzanna08} and \citet{tripathi09} have demonstrated that 
specific locations in an active region can show a change from
redshifts to blueshifts from the transition region to the
corona. \citet{delzanna09a,delzanna09b} presented velocity measurements in fan
loops near a sunspot and found evidence of decreasing redshift with
temperature: \ion{Fe}{vii} at $+30$~\kms, \ion{Fe}{viii} at
$+20$~\kms\ and \ion{Fe}{ix} at $+10$~\kms\ (lines formed at 0.6, 0.7
and 0.8~MK, respectively).
More recently, 
\citet{warren11} have demonstrated that the velocity structure within
the vicinity of
active region fan loops changes with temperature. They presented
velocity maps in different emission lines showing that the fan loops display
redshifts (downflows) for temperatures up to 0.8~MK. There is an
abrupt change to approximately zero velocity at a temperature of
1.0~MK, and then a change to strong blueshifts at higher temperatures
of 1.0--2.0~MK. These blueshifts represent the active region outflow
regions that have been described in earlier EIS papers
\citep{delzanna08,harra08,doschek08}.  
The authors suggested that the cool fan loops are distinct structures
from the active region outflows, with the latter probably representing
open field plasma that escapes to the solar wind, while the former are
closed loops.  Further support for a distinction between the outflow
regions and the fan loops is provided by \citet{ugarte11} who compared
time sequences of images and velocity maps from the EIS lines
\ion{Si}{vii} \lam275.37 ($T=0.7$~MK) and \ion{Fe}{xii} \lam195.12 (1.6~MK).

Inspection of the  images presented in \citet{warren11} shows that the
fan loops have a filamentary structure in both velocity and intensity
and thus there is the possibility that the bright (in intensity)
filaments within the loop may not necessarily correspond to filaments
with the strongest velocity shifts. 
The present work considers a specific example of a fan loop that is
distinct in intensity and that is aligned along the line-of-sight to
the observer (thus making the measured line-of-sight velocities a good
approximation to the actual loop velocities). The aim is to place the
qualitative observations of \citet{warren11} onto a quantitative
basis. A key part of the work is to make a general prescription for
deriving absolute velocities with correct error bars with EIS, which
has not been previously done.

\section{The EIS instrument and data reduction}\label{sect.eis}

Detailed descriptions of the EIS instrument are provided in
\citet{culhane07} and \citet{korendyke06}, and we summarize here only
properties that are important to the current investigation. EIS has a
single mirror for focusing and a single grating for dispersing the
solar spectrum. Both the mirror and grating are partitioned into two
halfs that have different multilayer coatings. The coatings are
optimized to yield high sensitivity in two wavelength bands of 170--212
and 246-292~\AA\ that we refer to as the short wavelength (SW) and
long wavelength (LW) bands, respectively. The two bands are imaged onto two distinct CCDs. The
image of the Sun returned by the mirror is sent through  one of a
choice of four slits. For the present investigation the data are
obtained with the narrowest slit, for which the width projected onto the
Sun is 1\arcsec. The exposure that is imaged
on the CCD contains spectral information along the CCD X-axis, and solar-Y
spatial information along the CCD Y-axis. A single pixel corresponds
to 22.3~m\AA\ in the spectral dimension and 1\arcsec\ in the spatial
dimension. To build up a 2D image of
the Sun, the mirror is tilted through consecutive steps of projected
size 1\arcsec\ after each
exposure. The maximum size of a single EIS raster is $540\times 512$~arcsec$^2$.

There are three instrumental issues that have a direct effect on
velocity measurements with EIS.
The most significant is an oscillation of
the grating due to varying thermal conditions during the 98~min orbit
of the Hinode satellite. It results in the positions of the spectra on
the detectors drifting in a quasi-sinusoidal manner by $\pm 1$~pixel
($\pm 23$--39~\kms, depending on wavelength). This orbital thermal
drift is typically corrected using the spectra themselves
\citep[e.g.,][]{mariska07}, but an independent method that makes use
of instrument temperature readings has recently been developed \citep{kamio10}.

The remaining issues that affect velocity measurements are related to
the EIS slits. Neither of the two narrow slits are exactly aligned
with the CCD axes, which means that the emission line
centroids  change with Y-position on the detector (an effect
usually referred to as ``slit tilt''). In addition the slits also show
a small but measurable curvature. \citet{kamio10} have accurately
measured the tilt and curvature from measurements of strong emission lines.

Two further instrumental issues affect the images obtained in
different emission lines.
The dispersion axis of the
grating is slightly tilted relative to the CCD X-axis which results
in the images of lines at different wavelengths being slightly offset
from each other \citep{young09b}. Secondly, the two CCDs are not exactly
aligned with each other which, when coupled with the grating tilt,
means that a solar feature that appears on a particular row on the LW detector
will appear between 15 and 20 pixels higher on the SW detector. Since
the same sets of rows are read out from each detector it means that
the field-of-view for the LW emission lines is different to the
field-of-view for the SW emission lines by 15--20\arcsec\ for each
exposure, thus a feature seen at the top or bottom of the image from
one channel may not be seen in the other channel.

The EIS data are calibrated with the \emph{Solarsoft} IDL routine
EIS\_PREP that is described in EIS Software Note No.~1 \citep{young11aa}, and
standard processing options were used in the present work. A
post-processing step was also performed whereby the intensity values
in CCD pixels flagged as
``warm'' or ``hot'' by EIS\_PREP were interpolated from neighboring
pixels using a method described in EIS Software Note No.~13
\citep{young10}. This method yields improved Gaussian fits to line
profiles compared to simply treating the warm and hot pixels as
missing data.

A list of the EIS emission lines studied in the present work are given
in Table~\ref{tbl.ions}, together with information about the
temperature at which the emission lines are formed. The temperature of
maximum ionization, $T_{\rm max}$, sometimes does not accurately describe the
temperature at which an emission line is principally emitted. If we
isolate the temperature dependent terms that contribute to a line's
observed intensity, $I$, we have
\begin{equation}
I \propto \int F(T)\epsilon(T,N_{\rm e})\phi(T){\rm d}T
\end{equation}
where $F(T)$ is the ionization fraction for the emitting ion,
$\epsilon(T,N_{\rm e})$ is the line emissivity computed by solving
the level balance equations for the ion, and $\phi$ is the
differential emission measure (DEM) curve. For spectral modeling, a
discretized temperature scale of fixed width in $\log\,T$ is often
used for computing the intensity, and 
so
\begin{equation}
I \propto \sum_i F_i\epsilon_i\phi_i T_i
\end{equation}
for a set of temperatures $T_i$ typically tabulated at fixed intervals
of $\log\,T$. When considering a particular DEM
curve, the temperature that yields 
the largest contribution to $I$ is the one for which the curve
$F(T)\epsilon(T)\phi(T) T$ has its maximum. We call this temperature the
\emph{effective temperature}, $T_{\rm eff}$, and the values for the ions
considered here assuming a quiet Sun DEM
curve are shown in Table~\ref{tbl.ions}. The calculation was performed
using version~6 of the CHIANTI atomic database \citep{dere97,dere09}
assuming a constant pressure atmosphere of $10^{14.5}$~K~cm$^{-3}$,
the ionization balance of \citet{dere09}, and the quiet Sun DEM that
is distributed with the CHIANTI database and which was computed from
the quiet Sun spectra of \citet{vernazza78}. The temperatures are
given to the nearest 0.05 in $\log\,T$.

Comparing the $T_{\rm max}$ and $T_{\rm eff}$ values in
Table~\ref{tbl.ions} the differences are mostly small, but for
\ion{Fe}{viii} the difference is very significant and arises from the
broad ionization fraction of this curve and the steeply rising DEM curve
in the upper transition region. This issue is discussed in detail by
\citet{brooks11}. The key point for the present paper is that
\ion{Fe}{viii} emission lines principally arise from the same plasma
that gives rise to \ion{Si}{vii} and \ion{Mg}{vii}. When referring to
temperatures of formation of the EIS lines in the remainder of this
paper we will use the $T_{\rm eff}$ values from Table~\ref{tbl.ions}.

\begin{deluxetable}{cccc}
\tablecaption{Emission lines used in the present work.\label{tbl.ions}}
\tablehead{
  \colhead{Ion} &
  \colhead{Wavelengths / \AA} &
  \colhead{$\log\,(T_{\rm max}/{\rm K})$\tablenotemark{a}} &
  \colhead{$\log\,(T_{\rm eff}/{\rm K})$}
}
\startdata
\ion{O}{iv} & 279.93 & 5.15 & 5.25\\
\ion{O}{vi} & 279.93 & 5.50 & 5.55 \\
\ion{Fe}{viii} & 185.21 & 5.60 & 5.85 \\
\ion{Mg}{vi} & 268.99 & 5.65 & 5.70\\
\ion{Mg}{vii} & 278.39, 280.72 & 5.80 & 5.85 \\
\ion{Si}{vii} & 275.37 & 5.80 & 5.85\\
\ion{Fe}{x} & 184.54 & 6.05 & 6.05 \\
\ion{Fe}{xi} & 188.22 & 6.15 & 6.10 \\
\ion{Fe}{xii} & 192.39 & 6.20 & 6.20 \\
\enddata
\tablenotetext{a}{Temperature of maximum ionization of the ion.}
\end{deluxetable}

\section{Overview of the data-set}

The aim of the present work is to make a definitive measurement of the
absolute velocity of the plasma in an active region fan loop by making
use of the wide temperature coverage of the EIS instrument.
Fan loops are found in virtually all active regions but for the present
work particular criteria were required: the footpoint regions must be
approximately orientated towards the observer's line-of-sight; the
footpoints must be isolated from other active region structures; and
there must be an area of quiet Sun within the raster field-of-view. The
first criterion ensures that the line-of-sight velocities measured
through emission line Doppler shifts are a close approximation to the
real motions within the loop structures. The second criterion is
required to prevent any contamination of the observed loop by
neighboring structures, and the third criterion ensures that a
velocity measurement relative to the quiet Sun can be made.

For this paper a single footpoint region was selected from active
region AR 11032 which crossed the solar disk during 2009 November
14--26. The region had emerged on October 22 during the
previous solar rotation when it produced several weak C-flares, but it
was quiescent during the second disk passage.
On November 21 23:44~UT an EIS raster covering the
west side of the region was obtained with the study
HPW015\_DETAILED\_MAP that was centered at heliocentric 
coordinates (+379,+251) in arcseconds. The raster scan used
45~sec exposures with the 1\arcsec\ slit for 200 steps, giving an
image of size 200\arcsec\ $\times$ 360\arcsec\ obtained over 2~hours 36~mins.

\begin{figure}[h]
\epsscale{1.0}
\plotone{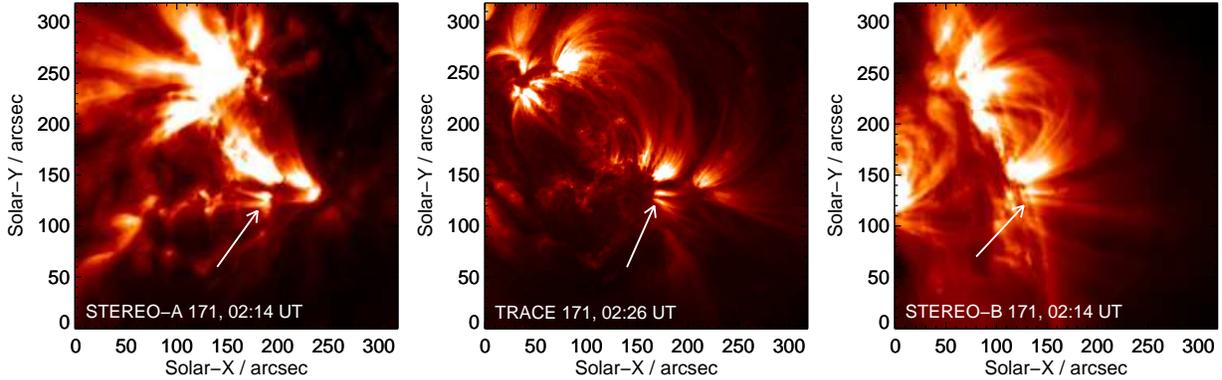}
\caption{Images of active region AR 11032 in the 171~\AA\ filters of the
  TRACE satellite (middle) and the EUVI instruments of the STEREO-A
  and STEREO-B spacecraft (left and right, respectively), obtained on
  2009 November~22. Both STEREO
  images were obtained at 02:14~UT and the TRACE image was obtained at
  02:26~UT. The locations of the two adjacent loop footpoints studied
  in the present work are indicated with an arrow in each image.}
\label{fig.euvi}
\end{figure}

The middle panel of Fig.~\ref{fig.euvi} shows the active region close to the
time of the EIS observation as seen in the 171 filter of the TRACE
satellite. This filter predominantly shows plasma at around one
million K and the fan loops are identified as structures that fan out
with height, sometimes connecting through large loops to the opposite
polarity side of the active region, or simply seen to extend out
from the active region until they can no longer be distinguished against
the background. In the latter case they may connect to another active
region, to the quiet Sun, or perhaps even extend out into the
heliosphere. Two loop footpoints are highlighted in the TRACE image of
Fig.~\ref{fig.euvi} that are bright and appear to have only a small
spatial extent. By considering the 171 filters of the A and B
STEREO spacecraft (left and right panels of Fig.~\ref{fig.euvi}, respectively) which
are at heliocentric angles of $+63.3^\circ$  and $-63.8^\circ$ compared to Earth, it is
apparent that the TRACE loops are quite extended and are aligned close
to the line-of-sight of the TRACE instrument. By using the 
angular separations of the two STEREO spacecraft and measuring the
projected lengths of the loops from the images shown in
Fig.~\ref{fig.euvi} we estimate through elementary geometry that the
angle of the loops to the TRACE 
(and thus \hinode) line-of-sight is around 20$^\circ$.
This means that the
loops are excellent candidates for measuring line-of-sight velocities
with EIS since the measured velocities will be close to the actual
velocities along the loops.

\begin{figure}[h]
\epsscale{1.0}
\plotone{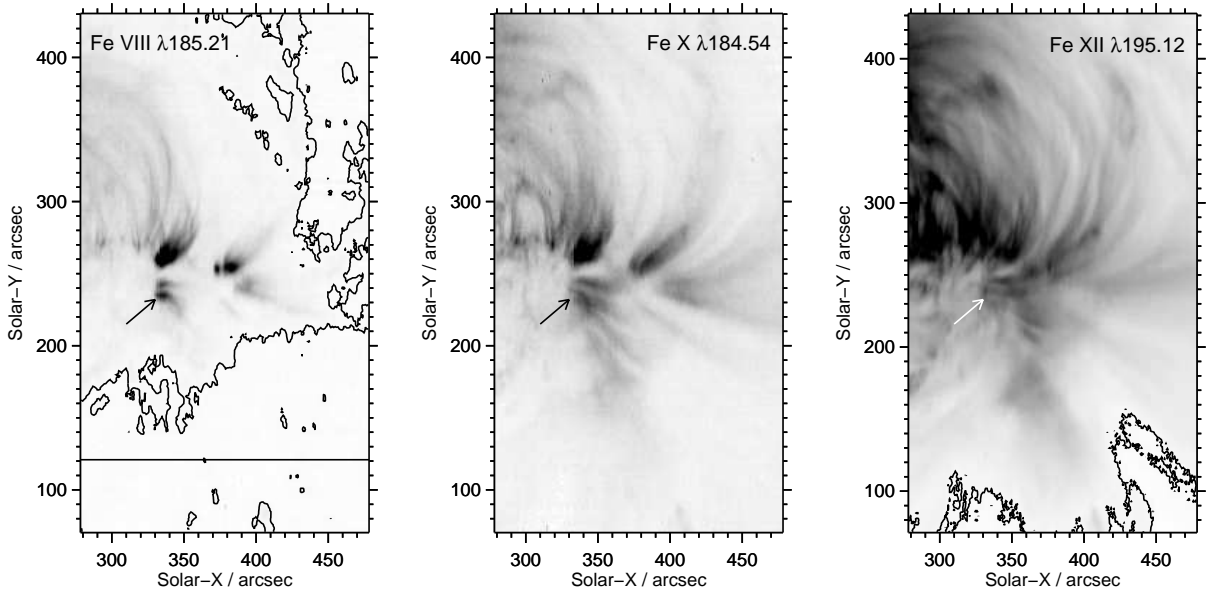}
\caption{Images from the \ion{Fe}{viii} \lam185.21, \ion{Fe}{x}
  \lam184.54 and \ion{Fe}{xii} \lam195.12 emission 
  lines derived from the EIS raster scan that began at 2009 November 21
  23:44~UT and lasted 2 hours 36 mins (EIS rasters from right to left). Black corresponds to regions of high intensity, and each 
  image has been saturated to reveal weaker emission regions. The
  contour in the left panel has been set at 1.5 times the average
  quiet Sun \lam185.21 intensity, and that in the right panel is set at
  two times the average
  quiet Sun \lam195.12 intensity (see text for more details). Arrows
  on each image point to the footpoint of the
loop that is analyzed in the present work. The horizontal line on the
\ion{Fe}{viii} image marks the upper boundary of the quiet Sun region
discussed in Sect.~\ref{sect.fe8}.}
\label{fig.eis-ar}
\end{figure}

Images of the active region from EIS are shown in
Fig.~\ref{fig.eis-ar}: the left panel shows an image formed from
\ion{Fe}{viii} \lam185.21 (formed around 0.7~MK); the middle panel
shows an image formed from \ion{Fe}{x} \lam184.54 (around 1.1~MK); and the right panel shows an image formed
from \ion{Fe}{xii} \lam195.12 (around 1.5~MK). The \ion{Fe}{x}
\lam184.54 image is most similar to the TRACE 171 image of
Figure~\ref{fig.euvi}. The \ion{Fe}{viii} \lam185.21 image shows a
number of compact, spiky structures that can be identified as the
footpoints of the \ion{Fe}{x} loops. The \ion{Fe}{xii} image shows the
same general loop structures as \ion{Fe}{x}, but the loops are more
extended and there appears to be many more of them.


\section{Measuring absolute velocities with EIS}\label{sect.abs-vel}

To determine absolute line-of-sight velocities it is necessary to be
able to compare the measured spectrum with a calibration spectrum
where the lines of interest are either at rest or at a known
velocity. Ideally this would be done using an on board  calibration lamp that
yields accurately known wavelengths. Unfortunately it is not practical
to fly calibration lamps for the  EIS wavelength ranges, and no
flight-quality lamp exists. There are several reasons for this,
including the lack of useful materials to use as the lamp window
(typical materials such as quartz, LiF and MgF are all opaque below
1000~\AA); large power requirements; and the weakness of the EUV spectrum
from gases that might be used in the lamp (e.g., Ne, Ar).

An alternative method is to use photospheric or low
chromosphere lines measured in the same spectrum as the lines of
interest. These cool lines typically show only very small velocity
shifts and thus can be treated as at rest, allowing the absolute
wavelength scale to be determined. An example of the use of this
method was provided by \citet{brekke97} who used emission lines from
neutral and singly-ionized species such as \ion{O}{i}, \ion{C}{i} and
\ion{Fe}{ii} to yield an absolute wavelength scale for the SOHO/SUMER
instrument. Justification for 
this method comes from the work of \citet{hassler91}, who
used data from a rocket-based spectrometer with an on board calibration
lamp to show that an emission line of \ion{Si}{ii} has a zero net
velocity shift in the quiet Sun. Additional support for the small
velocities of photospheric lines comes from the analysis of
\citet{samain91} who used absorption lines of \ion{Si}{i}, \ion{Fe}{i}
and \ion{Ge}{i} measured from a balloon experiment to determine an
absolute photospheric redshift of 
1~\kms. Although significant for the photosphere, this velocity is
small compared to typical transition region and coronal velocities and
means that the photospheric lines can be usefully considered to be at rest.

For the EIS wavelength ranges this method can not be used as the 
coolest emission line in the spectrum is
\ion{He}{ii} \lam256.32 (actually a self-blend of two lines) that is
formed around 80,000~K and so can not be 
 considered a photospheric or low chromosphere
line. The method of using photospheric or low chromosphere lines to set the absolute wavelength
scale can thus not be applied to the EIS spectra.


Another method for determining an absolute wavelength scale is to
assume that coronal lines have a zero net line-of-sight velocity above
the solar limb,  which is a plausible assumption given the long line-of-sight and the
likelihood that the line-of-sight components of any bulk plasma motions
would balance out. 
Two EIS exposures can then be made, one at the limb and one at the location
of the active region on the disk, with the limb spectrum yielding the
wavelength calibration.
This method does not work however due to the orbital thermal drift
effect discussed in Sect.~\ref{sect.eis}, which means that the spectral
position obtained at the 
limb will no longer be valid when the active region exposures are
taken on the disk. 

A further method for determining the absolute
wavelength calibration is to observe quiet Sun simultaneously with the
active region. That is, each individual EIS exposure will contain emission
from both active region and quiet Sun at locations along the slit. The
average absolute velocity of the quiet Sun as a function of
temperature was accurately established by \citet{peter99} using SUMER
spectra. Thus, for an EIS emission line formed at a specific
temperature, the quiet Sun velocity from the  velocity
curve shown in Fig.~6 of \citet{peter99} can be obtained and used to
convert the measured EIS quiet Sun 
velocity to a rest wavelength. This is the method used in the present
work.

A potential problem for this method is the highly dynamic nature of
the the quiet Sun  at transition region temperatures, which means that
a specific patch of quiet Sun may not show the velocity
pattern found by \citet{peter99}. Further, quiet Sun in the vicinity
of an active region may show significant differences from a quiet Sun region
far from centers of activity. These issues can not be investigated with the
EIS data, however we note that (i) the quiet Sun regions selected from the
EIS rasters are averaged over 50\arcsec\ in the Y-direction
(Fig.~\ref{fig.eis-ar}) and so
represent a spatial average over the quiet Sun; (ii) the error bars
provided by \citet{peter99} are derived from the spatial variation of
the velocity shifts over the quiet Sun and are included in the present
analysis; and (iii) the emission line
intensities found for \ion{Fe}{viii} in the quiet Sun region shown in
Fig.~\ref{fig.eis-ar} are consistent with an average quiet Sun value
(see Sect.~\ref{sect.fe8}). We thus believe it is a reasonable
approximation to apply the \citet{peter99} quiet Sun velocity results
to a quiet Sun location close to an active region.

An alternative means of deriving the absolute wavelength scale for EIS
data has been presented by \citet{kamio10}. Here an empirical
model that relates the variation
of the \ion{Fe}{xii} \lam195.12 line centroid to various temperature
measurements within the EIS instrument  over the entire EIS
mission has been created. By assuming that \ion{Fe}{xii} has a zero
net velocity shift for all EIS exposures, the model is able to yield
an absolute wavelength scale for any exposure obtained at any point in
the mission. As the temperature data used by the model comes from the
EIS housekeeping 
data-stream, this wavelength calibration method is referred to as the ``HK method''. \citet{kamio10} give
an accuracy for the HK method of 
4.4~\kms. A comparison between the HK method and the quiet Sun method used
here is presented in Appendix~\ref{app.kamio}, where it is seen that
there is a systematic offset in the two wavelength calibrations of
5.4~\kms. The quiet Sun method of
wavelength calibration is preferred here as it can be directly related
to previous, accurate measurements of solar velocities.


Even after the absolute wavelength scale has been determined for the
quiet part of the raster, the
instrumental issues mentioned in Sect.~\ref{sect.eis} mean that the
absolute wavelength scale can not be directly applied to the spatial
locations within the active region structures. Firstly, the EIS slits are
not straight nor are they exactly aligned to the axes of the
detectors. This means that the rest wavelength of a line will change
with the Y-location of the CCD, and so the rest wavelength determined
from the quiet Sun part of the slit image needs to be adjusted in
order to apply it to the active region structure. The tilts have been measured quite precisely by
\citet{kamio10} and the parameters from this work are used here.

A second instrumental issue that affects velocity measurements appears
to be due to a distorted point spread function for the EIS
instrument, and this is discussed in the following section.


\section{Asymmetric velocity distributions}\label{app.psf}

\begin{figure}[h]
\epsscale{1.0}
\plotone{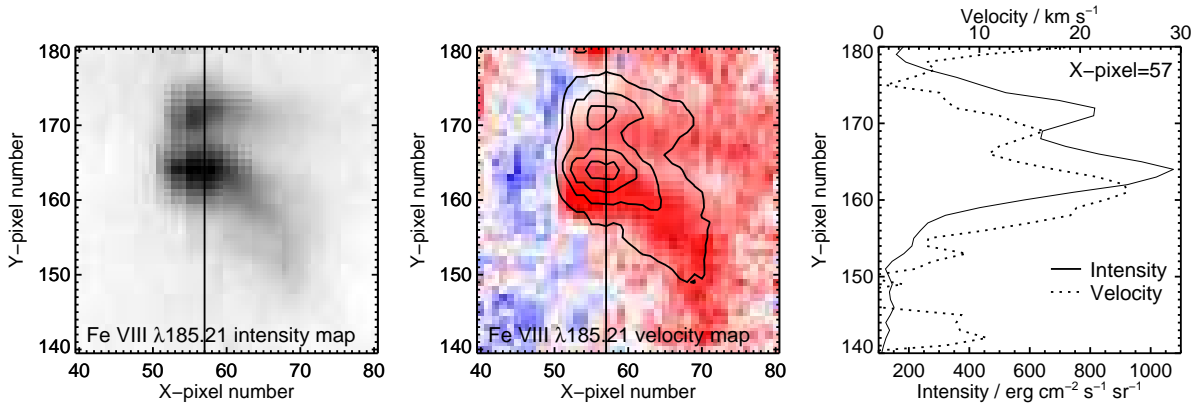}
\caption{The left panel shows the \ion{Fe}{viii} \lam185.21 intensity
  map in the vicinity of the loop footpoints. The middle panel shows
  the \lam185.21 velocity map of the same location with velocities
  between $\pm$~25~\kms\ displayed, and the contours
  of the intensity map overplotted. The right panel shows cuts through
the intensity and velocity maps at X-pixel 57, which is indicated on
the left and middle panels with a vertical, solid line. The solid and
dashed lines in the right panel show intensity and velocity,
respectively. Interpolation has been
peformed for missing data at X-pixels 47, 54 and 55.}
\label{fig.fe8vel}
\end{figure}

A striking feature of velocity maps obtained from EIS is that regions
of strongest redshift or blueshift are often spatially offset from
regions with the highest intensity. This effect is found in the
present work for the loop footpoints highlighted by an arrow in the
left panel of Fig.~\ref{fig.eis-ar} and is shown in
Fig.~\ref{fig.fe8vel}. A sub-region of size 41 pixels $\times$ 41
pixels has been extracted from the full \ion{Fe}{viii} image of
Fig.~\ref{fig.eis-ar} and Gaussian fits to the \lam185.21 line have been
performed at each spatial pixel to yield line intensity and centroid
measurements. Absolute velocities have been derived using the method
described in Sect.~\ref{sect.fe8}.
The slice through the data at X-pixel 57 shows
two distinct intensity peaks at Y-pixels 164 and 172, but the velocity
peaks occur at Y-pixels 161 and 169, respectively. Although this
feature could be explained by plasma rotating around the axes of the
loops, a survey of several loop footpoints performed for this work
showed that, where significant redshifts could be identified from the
\ion{Fe}{viii} line, they were always to the south of the location
where the intensity peaked, no matter where the active region was
located on the Sun. In
addition, it is clear from inspection of high resolution TRACE images
that loops that are apparently monolithic actually comprise multiple,
narrow features thus a large scale twisting flow is difficult to
interpret within this type of physical structure.

To state the observed effect simply,
wherever there is a \emph{decreasing} intensity gradient from north to
south, the centroid of the emission line will be artifically shifted
to longer wavelengths (redshift); and wherever there is an
\emph{increasing} intensity gradient from north to
south, the centroid of the emission line will be artifically shifted
to shorter wavelengths (blueshift).
Observations of polar coronal holes provide another illustration of
the effect that is apparent due to 
limb brightening in coronal lines. \citet{tian10} presented
velocity maps of the north polar hole obtained with EIS where a
distinctive ridge of redshifts is found along the limb in the
\ion{Fe}{xii} \lam195.12 and \ion{Fe}{xiii} \lam202.04 emission
lines. This arises because there is a decreasing intensity gradient
from north to south at the limb. Inspection of velocity maps from the
south polar hole show a ridge of blueshifts as the intensity gradient
is in the opposite direction. \citet{tian10} also reported that the
bright points within the coronal hole systematically displayed
blueshifts on one side and redshifts on the other. The authors
interpreted this as siphon flows within the structures but it may
actually be due to the strong intensity gradients when bright points are
observed within a dark coronal hole.

A similar effect was found from spectra from the Coronal Diagnostic
Spectrometer \citep[CDS;][]{harrison95} instrument on board SOHO
and was explained in terms of an asymmetric point spread
function (PSF) by \citet{haugan99}. Consider a PSF in the shape of a
2-dimensional Gaussian function, but with an elliptical
cross-section. The axes of the ellipse are tilted relative to the
wavelength--solar-Y axes of the detector such that the long axis of
the ellipse is at an
angle of 135$^\circ$ to the wavelength axis. Now if a bright spot of
emission is imaged, the photon counts will appear on the detector as
an ellipse. On the north side of the ellipse, there is an excess of
counts to the short wavelength side of the spot, while on the south
side of the ellipse there is an excess of counts on the long
wavelength side of the spot. Measuring the centroid of the spot as a
function of solar-Y will thus show a blueshift on the north side of
the spot, and a redshift on the south side, consistent with the
pattern shown in Fig.~\ref{fig.fe8vel}. Note that if the PSF was
rotated through 90$^\circ$ then the opposite velocity pattern would
appear in the observations.


The PSF will be expected to vary with position along the 1024\arcsec\
length of the EIS slit due to optical properties of the spectrometer, therefore the pattern of redshifts and
blueshifts induced by the PSF would be expected to change, but this is
beyond the scope of the present work. 
If the above interpretation of the data is correct, then the simplest
way of deriving velocity shifts for 
features with steep intensity gradients is either to study only those pixels
at the location of the intensity peaks in the solar-Y direction, or to
perform averaging in the Y-direction over a region symmetrically
distributed around the intensity peak. This is the procedure employed
in the present work.

\section{Velocity derivation method}

In this section we describe the particular steps involved in deriving the
absolute velocities and associated uncertainties for the 2009 November
21 data-set. As described in Sect.~\ref{sect.abs-vel}, the method requires
a determination of the rest wavelength scale from a patch of quiet Sun
that is observed simultaneously with the active region. Ideally the
strongest line in the spectrum should be used for this in order to
give the best possible measurement in the quiet Sun. However the
strongest EIS lines from the quiet Sun are all formed at temperatures
of 1--2~MK, the temperature range where the
active region has its largest spatial extent. This is illustrated in
the right panel of Fig.~\ref{fig.eis-ar} which shows the raster image
obtained in \ion{Fe}{xii} \lam195.12, the strongest line observed by
EIS (formed at 1.5~MK). Bright loop structures are clearly seen to
extend all the way to the top of the raster and, although the bottom
of the raster appears to be dark (and thus potentially classed as
quiet Sun), the intensity is still significantly brighter than average
quiet Sun levels. This is indicated by the black contour which shows
locations where the \lam195.12 intensity is a factor 2 brighter than
the average quiet Sun value measured by \citet{brooks09}. At no point in the \lam195.12
image does the intensity fall to average quiet Sun levels.

The \ion{Fe}{viii} \lam185.21 image shown in the left panel of
Fig.~\ref{fig.eis-ar} shows that the active region has a significantly
smaller spatial extent at cooler temperatures. The contour, which
in this case is set at 1.5 times the quiet Sun \lam185.21 intensity
measured by \citet{brooks09}, shows that a significant fraction of the
image at the bottom of the raster can be considered to be quiet
Sun. For this reason we choose to use \ion{Fe}{viii} \lam185.21 as the
line from which the rest wavelength scale is determined in the present
work.

\ion{Fe}{viii} represents the ideal ion to use for the EIS wavelength
calibration as (i) it is sufficiently strong that it can be measured
accurately in the quiet Sun, and (ii) the active region has a relatively
small spatial extent in this ion. Emission lines from cooler ions in
the EIS wavelength ranges, such as \ion{O}{vi} \lam184.11
and \ion{Mg}{vi} \lam268.99, are too weak to be reliably measured in the quiet
Sun, while the active region is too extended in images from the hotter
coronal ions (\ion{Fe}{x--xiii}) to enable the quiet Sun to be
seen within the EIS field of view. The isolated \ion{Fe}{ix} \lam197.86 line \citep[first
identified by][]{young09a} is possibly an alternative to
\ion{Fe}{viii} but was not measured in the present
data-set. \ion{Si}{vii} \lam275.37 is formed at a similar temperature
to \ion{Fe}{viii} \lam185.21 (Table~\ref{tbl.ions}) but is a factor two
weaker and so not as easily measured in the quiet Sun.
\lam186.60 and \lam194.66 are two other \ion{Fe}{viii} lines that
could have been used as references since their signal-to-noise is
comparable to \lam185.21 \citep{young07b,brown08}. However, \lam186.60 was not observed in the
2009 November 21 raster while \lam194.66 is partially blended with a
line in the long wavelength wing and so is less suitable than
\lam185.21 for velocity measurements.

\ion{Fe}{viii} \lam185.21 is used to determine a rest wavelength from
the quiet Sun and the detailed method for doing this is described in
Sect.~\ref{sect.fe8}. For other lines it is necessary to calibrate
relative to
\lam185.21 by making use of standard wavelength separations that are
measured from the EIS spectra. Different methods are required for the
two EIS wavelength channels and these are described in
Sects.~\ref{sect.sw} and \ref{sect.lw}, respectively.

\subsection{Velocities from Fe\,VIII \lam185.21}\label{sect.fe8}

\begin{figure}[h]
\epsscale{0.7}
\plotone{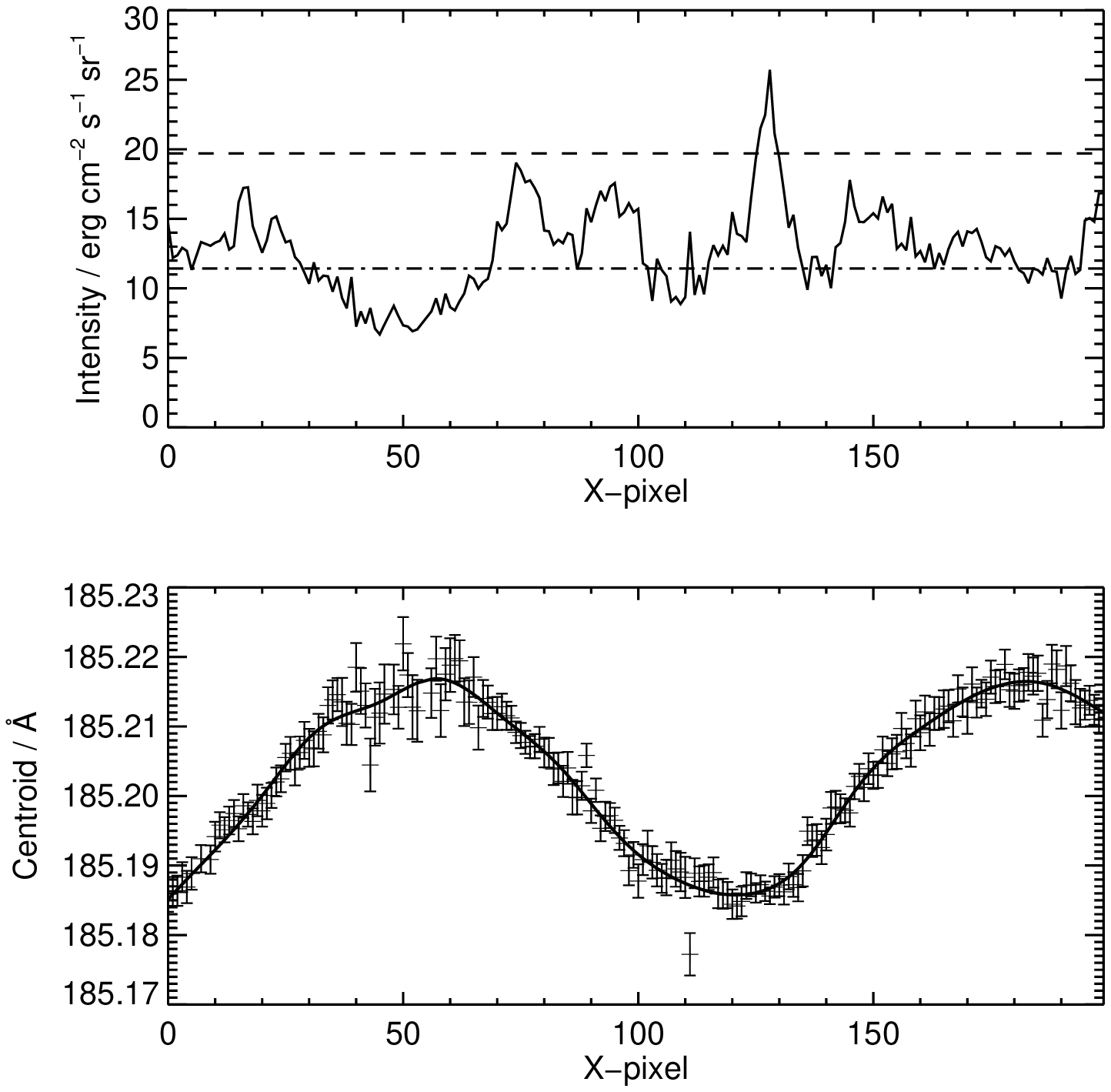}
\caption{The upper panel shows the variation of the \ion{Fe}{viii}
  \lam185.21 emission line intensity in the quiet Sun band of the 2009
November 21 raster. The intensity at each position is derived from an
average of 50 spatial pixels in the solar-Y direction. The dashed line
denotes the average quiet Sun intensity of \ion{Fe}{viii} \lam185.21
found by \citet{brooks09}, and the dash-dot line shows this value
multiplied by 0.58 (see main text for more details). The lower panel
shows the measured \lam185.21 line centroids at each of the quiet Sun
pixels, demonstrating the orbital variation of line position. The
thick solid line shows the spline fit to the centroids.}
\label{fig.fe8-qs}
\end{figure}

The first step is to fit the \ion{Fe}{viii} \lam185.21 line in the quiet Sun region
at the bottom of the raster (see the left panel of Fig.~\ref{fig.eis-ar}).
The bottom 50 Y-pixels of the raster image
are selected and the data are binned in the
Y-direction over this 50 pixel region, then Gaussian fits performed to
each of the 200 pixels in the X-direction. 
Binning is required to give sufficient signal in the \lam185.21 line
to allow the line centroid to be measured accurately; it also ensures
that any local inhomogeneities in the quiet Sun are smoothed over.
The \lam185.21 intensity over
the 200 X-pixels is shown in the upper panel of Fig.~\ref{fig.fe8-qs}
where the level is mostly below the average quiet Sun intensity found
by \citet{brooks09}. We note that the EIS sensitivity has
probably decreased between the present observation and that of
\citet{brooks09}, which was obtained on 2007 January 30. Preliminary
measurements of the sensitivity decay (J.T.~Mariska, 2011, private
communication) suggest an exponential decay with a 1/e time of 1894
days. This would imply that the EIS sensitivity has decreased by a
factor 0.58 between the two observations. The dash-dot line in
Fig.~\ref{fig.fe8-qs} shows the \citet{brooks09} quiet Sun measurement
multiplied by 0.58, placing it in good agreement with the intensities
measured in the quiet part of the \ion{Fe}{viii} raster. This confirms
that the quiet region considered here can be reasonably
classed as quiet Sun.  The bottom panel of Fig.~\ref{fig.fe8-qs} shows
the variation of the \lam185.21 centroid over the quiet Sun region.
The oscillatory pattern reveals the orbital thermal drift described
in Sect.~\ref{sect.eis}. A spline fit to the \lam185.21 centroid 
variation is then performed, with the spline defined at 10~minute
nodes (recall that each X-pixel represents a different point in time
due to the nature of the EIS observations; 10~mins corresponds to
around 13 X-pixels for this data-set). 
The standard deviation of the difference between the actual centroid measurements and the fit is 2.1~m\AA. This is the uncertainty in the quiet Sun wavelength scale, denoted by $\sigma_{\rm qs}$,
 which is used in the error analysis later.  

\ion{Fe}{viii} \lam185.21 will not be at rest in the quiet Sun and it
is necessary to determine the average quiet Sun velocity using the
velocity curve provided by \citet{peter99}. This is done by finding
which of the ions used by \citet{peter99} is closest in temperature to
\ion{Fe}{viii}. Table~\ref{tbl.ions} gives a $T_{\rm eff}$ value of
5.85 which fortuitously matches the value for \ion{Ne}{viii} \lam770.41, one of
the lines used by \citet{peter99}.
The average velocity shift for this
line is $v_{\rm PJ} =-2.6\pm 1.8$~\kms\ (blueshift), and therefore we assume this applies to
\ion{Fe}{viii} \lam185.21.\footnote{Note that \citet{peter99} used
  positive velocities to denote blueshifts whereas we use the more
  standard notation of positive numbers denoting redshifts.} If we take
the measured quiet Sun wavelength of \lam185.21 as determining the
rest wavelength at Y-pixel $q$, the center of the chosen quiet Sun
region (pixel 25 in the present case), then the rest wavelength for a
X-pixel $i$ is:
\begin{equation}\label{eq.rest}
\lambda_{\rm rest}(i,q)=\lambda_{\rm qs}(i,q)- {v_{\rm PJ} \over c} \lambda_{\rm qs}(i,q) \cos\,\theta
\end{equation}
where $\lambda_{\rm qs}$ is the wavelength from the spline fit shown
in the bottom panel of Fig.~\ref{fig.fe8-qs}, $c$ is the speed of
light, and 
$\theta$ is the heliocentric angle ($\theta=0$ is disk center). For
the region considered here, $\cos\,\theta=0.90$--0.93 
and, for simplicity, we take a value of 0.92 at all spatial pixels in
the raster. 
The rest wavelength will change with Y position due to the EIS
slit tilt such that
\begin{equation}
\lambda_{\rm rest}(i,j)=\lambda_{\rm rest}(i,q)+t(j)-t(q)
\end{equation}
The EIS slit tilts, $t(j)$, are described by cubic polynomials and the
coefficients are given by \citet{kamio10}.

If we consider a pixel $(i,j)$ within the active region loop, then we
now have the rest wavelength at this pixel position against which the
loop velocity can be determined. Since velocities at individual pixels
are affected by the point spread function problem
(Sect.~\ref{app.psf}), however, then  we choose for the present
study to
average spectra from the loop over Y-regions that span the loop
cross-section; the asymmetries  then cancel out to yield more
accurate velocities. The Y-regions selected for each X-pixel location
are discussed in Sect.~\ref{sect.loop}.
The measured loop velocities thus
correspond to pixels $(i,p_i)$ where $p_i$ is the Y-pixel corresponding to
the center of the chosen loop Y-region at each X-pixel $i$.

The error on the absolute velocity for \ion{Fe}{viii} \lam185.21
comprises four components that are added in quadrature:
\begin{equation}\label{eq.fe8-error}
\sigma_{\lambda}^2 = \sigma_{\rm loop}^2 + \sigma_{\rm tilt}^2 + \sigma_{\rm qs}^2
                  + \sigma_{\rm PJ}^2
\end{equation}
where $\sigma_{\rm loop}$ is the measurement error of the emission line
wavelength in the loop, $\sigma_{\rm tilt}$ is the uncertainty in the
determination of the slit tilt parameters, $\sigma_{\rm qs}$ is the
uncertainty in the determination of the orbit variation, and
$\sigma_{\rm PJ}$ is the uncertainty in the absolute velocity of
\ion{Fe}{viii} \lam185.21 in the quiet Sun as obtained from
\citet{peter99}. For the present case the dominant  error term
is $\sigma_{\rm qs}$ which is 2.1~m\AA; $\sigma_{\rm PJ}$ is
1.1~m\AA, $\sigma_{\rm tilt}$ is between 0.1 and 0.2~m\AA, and
$\sigma_{\rm loop}$ varies from 0.4~m\AA\ to 1.4~m\AA\ depending on
the signal in the loop line profile. Note that $\sigma_{\rm
  loop}$ and $\sigma_{\rm qs}$ are random errors, whereas
$\sigma_{\rm tilt}$ and $\sigma_{\rm PJ}$ are systematic errors. The
combined errors lead to absolute velocity uncertainties of 4--5~\kms\
for \ion{Fe}{viii} \lam185.21.

\subsection{Velocities for other SW lines}\label{sect.sw}

For other emission lines within the EIS SW band, the above limb
wavelength offsets of \citet{warren11} are used:  the
\ion{Fe}{viii} \lam185.21 rest
wavelength at loop position $(i,p_i)$ sets the absolute wavelength
scale and the rest wavelengths of other lines are obtained from the
wavelength offsets relative to \lam185.21
given in Table~1 of \citet{warren11}. The wavelength offsets have
uncertainties, $\sigma_{\rm limb}$, that are obtained from
\citet{warren11} by combining in quadrature the error 
for the line of interest with that of \lam185.21. \citet{warren11} did
not include \ion{O}{vi} \lam184.11 although this line can be seen in
off-limb spectra. Using the same data-set as \citet{warren11} we find
a wavelength of $184.1161\pm 0.0005$~\AA.

\subsection{Velocities in the LW band}\label{sect.lw}

Potentially the method of using off-limb wavelength offsets can be
extended to the EIS LW band. However, studying a number of quiet Sun
data-sets it is apparent that the  offset of \ion{Si}{vii}
\lam275.35 relative to \lam185.21 shows variations of up to
0.02~\AA\ over long time-scales. The procedure for checking this was
to consider quiet Sun rasters and measure the \ion{Fe}{viii} and
\ion{Si}{vii} lines in ``macro-pixels'' of size 2$\times$20 pixels
(for example) that are chosen to yield strong enough signals for
measuring line centroids accurately. Constructing histograms of the
wavelength differences between the lines across the raster yields a Gaussian-shaped
curve with a full-width at half maximum of around 2--3~m\AA. However,
the centroid of the Gaussian distribution varies from 90.16 to
90.18~\AA. The small dispersion in the measured separations is
consistent with the two emission lines having the same $T_{\rm eff}$
value (Table~\ref{tbl.ions}), but the longer
time-scale variation is a surprise and suggests that slow thermal
effects within the instrument may cause the wavelength scale to
stretch or shrink over time. The changes are small for lines within
the SW band due to the small wavelength separations, but become
significant when comparing SW to LW separations.
The end result is that SW--LW separations from an off-limb data-set such as that analyzed
by \citet{warren11} can not be applied to a data-set taken at a
different time.

An alternative is to treat the LW channel independently, and measure
\ion{Si}{vii} \lam275.37 in the quiet Sun part of the active region
raster in the same manner as was done for \ion{Fe}{viii} \lam185.21
However, \lam275.37 is around a factor two weaker than
\ion{Fe}{viii} \lam185.21 and, coupled with the fact that around 18 of
50 quiet Sun pixels are lost because of the offset between the two EIS
CCDs (Sect.~\ref{sect.eis}), the \lam275.37
signal in the quiet Sun region is insufficient to reliably obtain
the orbit variation. 

The solution for the present work is to assume that the velocity shift
of \ion{Si}{vii} \lam275.37 at a specified spatial location in the
loop is the same 
as that of \ion{Fe}{viii} \lam185.21. This is suggested by the fact
that the images in the two lines are so similar, and the assumption can be
checked using the present data-set. 

\begin{figure}[h]
\epsscale{0.7}
\plotone{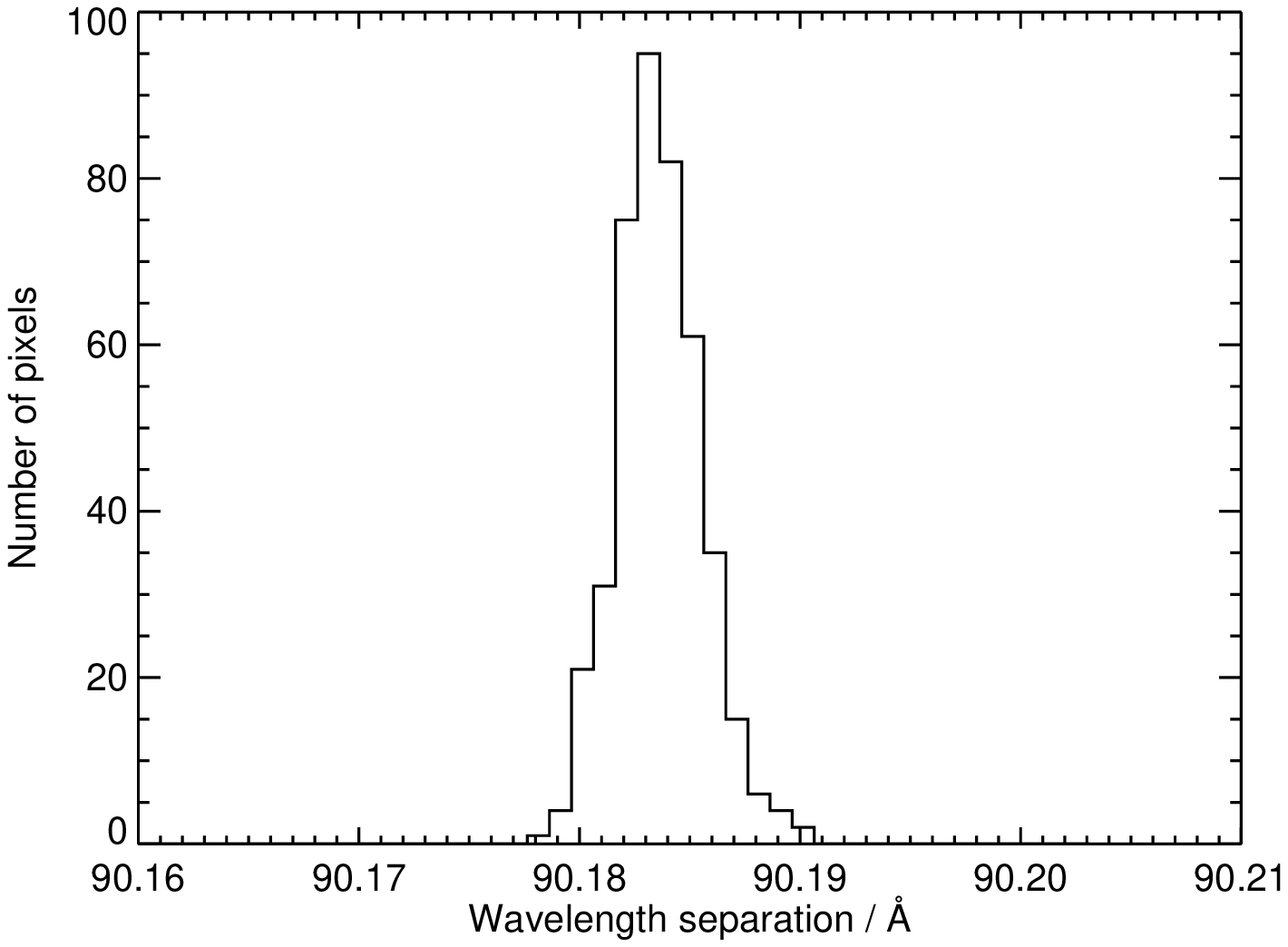}
\caption{Histogram showing the wavelength separations of the
  \ion{Si}{vii} \lam275.37 and \ion{Fe}{viii} \lam185.21 lines over
  432 pixels in the loop footpoint regions of AR 11032.}
\label{fig.fe8-si7}
\end{figure}

If the two lines do show the same velocity shift, then if we create a
histogram of the difference in measured wavelength between the two
lines, we would expect to see only a small scatter. The two lines are
normally too weak to extract reliable centroid measurements in
individual spatial pixels, but they become sufficiently bright in the
loop footpoints that this is possible. We filter out all spatial
pixels in the November 21 raster that have a \ion{Si}{vii} \lam275.35
intensity above 300~\ecs\ (437 in all). We then remove those
pixels for which \lam185.21 is not reliably measured, giving 432
pixels. (Note that a CCD spatial offset of 18 pixels between the two
emission lines needs to be accounted for when doing
this comparison.)
For these 432 pixels, Fig.~\ref{fig.fe8-si7} shows
the distribution of wavelength separations. The mean separation is
90.1841~\AA, but more importantly the standard deviation is only
0.0020~\AA, corresponding to a velocity shift of 2.1~\kms. 
Part of this dispersion will be due to the different slit tilts that
apply to the two wavebands. All of the selected spatial pixels lie within a 50 pixel
band in the Y-direction, over which the slit tilt difference changes
by only 1~m\AA, so this is only a minor contribution.
The small dispersion in the \lam275.37--\lam185.21 wavelength
separation gives 
confidence that the \ion{Si}{vii} \lam275.37 velocity shifts mimic
those of \ion{Fe}{viii} \lam185.21, and so supports our assumption
that \lam275.37 has the same velocity shift as \lam185.21.

The rest wavelength of \lam275.37 within the loop is thus obtained by
taking the measured \lam185.21 loop velocity and correcting the
measured \lam275.37 wavelength in the loop by this velocity. For
example, suppose
\lam185.21 has a blueshift of $15$~\kms\ in the loop and, at the same
spatial location, the measured wavelength of \ion{Si}{vii} \lam275.37
is 275.355~\AA. The \ion{Si}{vii} line is assumed to have the same
velocity shift as the \ion{Fe}{viii} line, so adding a shift of
+15~\kms\ to the measured \ion{Si}{vii} \lam275.37 wavelength yields
the rest wavelength of 275.369~\AA.
An
additional error component of 0.0020~\AA\ is added in quadrature to the \lam185.21
error components (Eq.~\ref{eq.fe8-error}) to account for the
dispersion in Fig.~\ref{fig.fe8-si7}.

For other lines in the LW channel the off-limb wavelength offsets of
\citet{warren11} can be used to yield the rest wavelengths from the
\lam275.37 rest wavelength. However, our primary interest in the LW
channel for the present work is to give access to the cool emission
lines \ion{O}{iv} \lam279.93 and \ion{Mg}{vi} \lam268.99 that
can not be reliably measured in off-limb spectra. The \ion{O}{iv} line
is particularly interesting as it is predicted to be formed at
180,000~K in the quiet Sun (Table~\ref{tbl.ions}) and represents the coolest emission line
that can be used for velocity measurements with EIS. 

The procedure for determining rest wavelengths for \ion{O}{iv} and
\ion{Mg}{vi} from the \lam275.37 line is as follows: all three lines
were measured in a set of eight average quiet Sun spectra obtained with
EIS in 2007 (six spectra), 2009 and 2010 (one spectrum each). The spectra
were averaged over a sufficiently large
spatial region to yield good measurements of the line centroids.
The
separations of the lines were computed, and an average taken. An error
on the separation was taken from the standard deviation of the
different quiet Sun measurements. The offsets relative to \lam275.37
are found to be 
$4.5724\pm 0.0025$~\AA\ and $-6.3816\pm 0.0012$~\AA\ for \ion{O}{iv}
\lam279.93 and \ion{Mg}{vi} \lam268.99, respectively.
Now these offsets apply to the quiet Sun, which has its own velocity
structure. Therefore the separations need to be corrected for the
average quiet Sun velocities of \citet{peter99} to yield the rest
wavelength separations. \ion{O}{iv} was measured by \citet{peter99} in
the quiet Sun and was found to have a velocity shift of $+9.9\pm
2.7$~\kms. \ion{Mg}{vi} was not measured by \citet{peter99}, but from
the CHIANTI atomic data we find that it is expected to be formed at
$\log\,T=5.7$ in the quiet Sun placing it  between \ion{O}{vi}
and \ion{Ne}{viii}, which were measured by \citet{peter99}. We assign
\ion{Mg}{vi} a quiet Sun velocity of $0.0\pm 2.0$~\kms. For
\ion{Si}{vii}, we assume it has the same quiet Sun velocity as
\ion{Fe}{viii}  and \ion{Ne}{viii}. 

The rest wavelength separations of  \ion{O}{iv} to \ion{Si}{vii},
and \ion{Mg}{vi} to \ion{Si}{vii} are thus $4.5607\pm 0.0039$~\AA\
and $-6.3840\pm 0.0027$~\AA, respectively. These values are used to
determine the \ion{O}{iv} and \ion{Mg}{vi} rest wavelengths from the
\ion{Si}{vii} rest wavelength.

\section{Large-scale velocity structure}\label{sect.loop}

Figures~\ref{fig.euvi} (middle panel) and \ref{fig.eis-ar} identify
two nearby loop footpoints in AR 11032 that are 
well defined in
the TRACE 171 channel and EIS \ion{Fe}{viii} \lam185.21. The aim of
the present work is to measure the line-of-sight velocities along one of
these loops from ion emission lines formed at different
temperatures. For this it is necessary to have confidence that the 
loop is well-defined at each temperature considered. For ions formed
at temperatures similar to \ion{Fe}{viii} or below (see
Table~\ref{tbl.ions}) this is straightforward as the images formed in
these ions show the same pair of compact brightenings that the \ion{Fe}{viii}
image shows. For higher temperatures the loop identification becomes
more difficult. 

\begin{figure}[t]
\epsscale{1.0}
\plotone{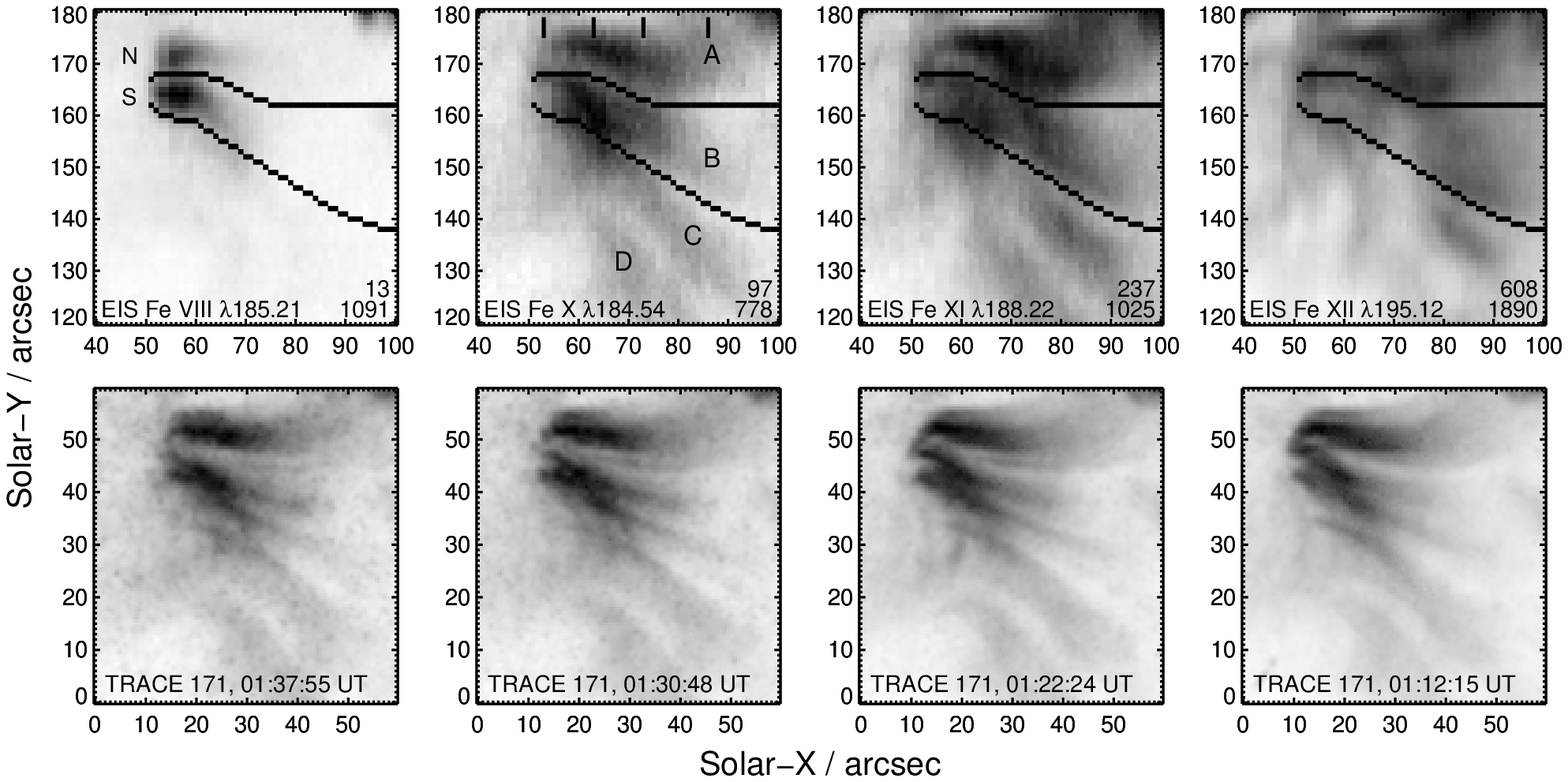}
\caption{The upper four panels show the loop footpoint region in four
  emission lines observed by EIS, which scanned from right to left 
  between 01:01 and 01:48~UT on 2009 November 22.  The four lower panels show the same footpoint region
  observed in the TRACE 171 filter at four different times during the EIS
  raster. The EIS spatial locations corresponding to these times are
  indicated by the thick, black lines near the top of the EIS
  \ion{Fe}{x} image.
  The two black lines in each of the EIS images identify the loop
  structure analyzed in the present work.
The numbers in the bottom
  right of each EIS image give the minimum (upper) and maximum (lower)
  intensity values (units: \ecss) in each image.
A linear, inverse scaling is used for all images,
  with black indicating regions of high intensity. The X and Y-axes
  for the EIS images are relative to the bottom-left corner of the
  complete EIS raster; those for the TRACE image are relative to the
  bottom-left corner of the displayed images.
}
\label{fig.eis-ims}
\end{figure}

The four upper panels of Fig.~\ref{fig.eis-ims} show intensity images
from the lines \ion{Fe}{viii} \lam185.21, \ion{Fe}{x} 
\lam184.54, \ion{Fe}{xi} \lam188.22 and \ion{Fe}{xii} \lam195.12 for a
61~pixel~$\times$~61~pixel sub-region of the full EIS raster that
shows the footpoint region. The two distinct footpoints are clearly
seen in the \ion{Fe}{viii} \lam185.21 image where they are labeled `N' and `S' for north and south,
respectively. The \ion{Fe}{x} \lam184.54 image shows much more extended
structures, and two loops labeled `A' and `B' can be matched with
the two footpoints N and S. There are two additional patches of
enhanced emission labeled `C' and `D' that may be loops that connect
to footpoint S. The higher spatial resolution of the TRACE 171 channel
images can be used to study the morphology of the EIS \ion{Fe}{x}
image in more detail as the principal component to the channel's
response comes from \ion{Fe}{x}. The four lower panels of
Fig.~\ref{fig.eis-ims}  show TRACE 171 images obtained at four times
during the course of the EIS raster. The times at which the TRACE
images were taken are indicated by four short vertical lines at the
top of the
EIS \ion{Fe}{x} image. The TRACE images show that the basic
morphology of the loops remains the same during the EIS
raster. Footpoint S appears to have two narrow, nearby 
loops associated with it, although these are not resolved in the EIS
data. The upper of these narrow loops appears to connect to the solar
surface between S and N at the early part of the raster, but is closer
to S when EIS rasters the footpoint. The upper loop structure A is
clearly separated from B and connects to footpoint N. In the TRACE
images, the structures C
and D do not connect to footpoint S.

Loop A is more distinct than loop B, however the key \ion{Fe}{viii}
\lam185.21 line is affected  by missing pixels  in the center of
the line profile that would affect velocity determinations. We therefore
 focus on loop B and the footpoint S, which are identified by
a pair of 
black lines in each of the upper panels of Fig.~\ref{fig.eis-ims} that
were drawn based on a study of
the EIS \ion{Fe}{viii} and \ion{Fe}{x} images, and the TRACE 171
images. Since the TRACE 171 images show that loop B consists of two
distinct strands then we will generally refer to it as ``loop
structure B'' to denote that it is not a single monolithic
structure. Indeed the discussion presented in the following paragraphs
suggests there is additional structure to that seen in the TRACE 171
images. The common feature of the individual strands within loop
structure B is that they appear to terminate in footpoint S.




Moving to the hotter \ion{Fe}{xi} \lam188.22 line, the structures A--D become
brighter away from the footpoint region but also the emission, in
general, becomes more diffuse over the raster. Considering structure B
there appears to be two brightness components: one that spatially
matches the bright region of \ion{Fe}{x} (around X-pixels 55--70), and
another that extends to larger heights (X-pixels 70--90). The two components are seen
more clearly in the \ion{Fe}{xii} \lam195.12 image where there is a definite gap
between the extended emission around X-pixels 75--100 and the  \ion{Fe}{x}
brightness patch around X-pixels 55--70. In general, the \ion{Fe}{xii}
image shows little relation to the very simple two footpoint image of
\ion{Fe}{viii} and thus hints at increasing complexity with
height/temperature.

Our interpretation of the image data is that there are many loops
coming from the both the \ion{Fe}{viii} bright footpoints (N and S)
and other footpoints that are not bright in \ion{Fe}{viii} (e.g., the
invisible footpoints of structures C and D). These loops have a range
of angles to the solar surface, perhaps related to where the loops
connect back to the solar surface on the opposite side of the active
region (see below). For low temperatures/height (\ion{Fe}{viii}) the different loops
converge and give the impression of distinct, bright patches of
emission. For larger temperatures/height (\ion{Fe}{xii}) the loops
diverge from each other and give much less distinct emission
signatures. In particular, the \ion{Fe}{x} brightness patch of loop B
that is seen in both \ion{Fe}{xi} and \ion{Fe}{xii} is likely a
distinct structure from the extended, diffuse emission seen in
\ion{Fe}{xi} and \ion{Fe}{xii} over X-pixels 75--100, even though both
have their footpoints within footpoint S of \ion{Fe}{viii}.

\begin{figure}[h]
\plotone{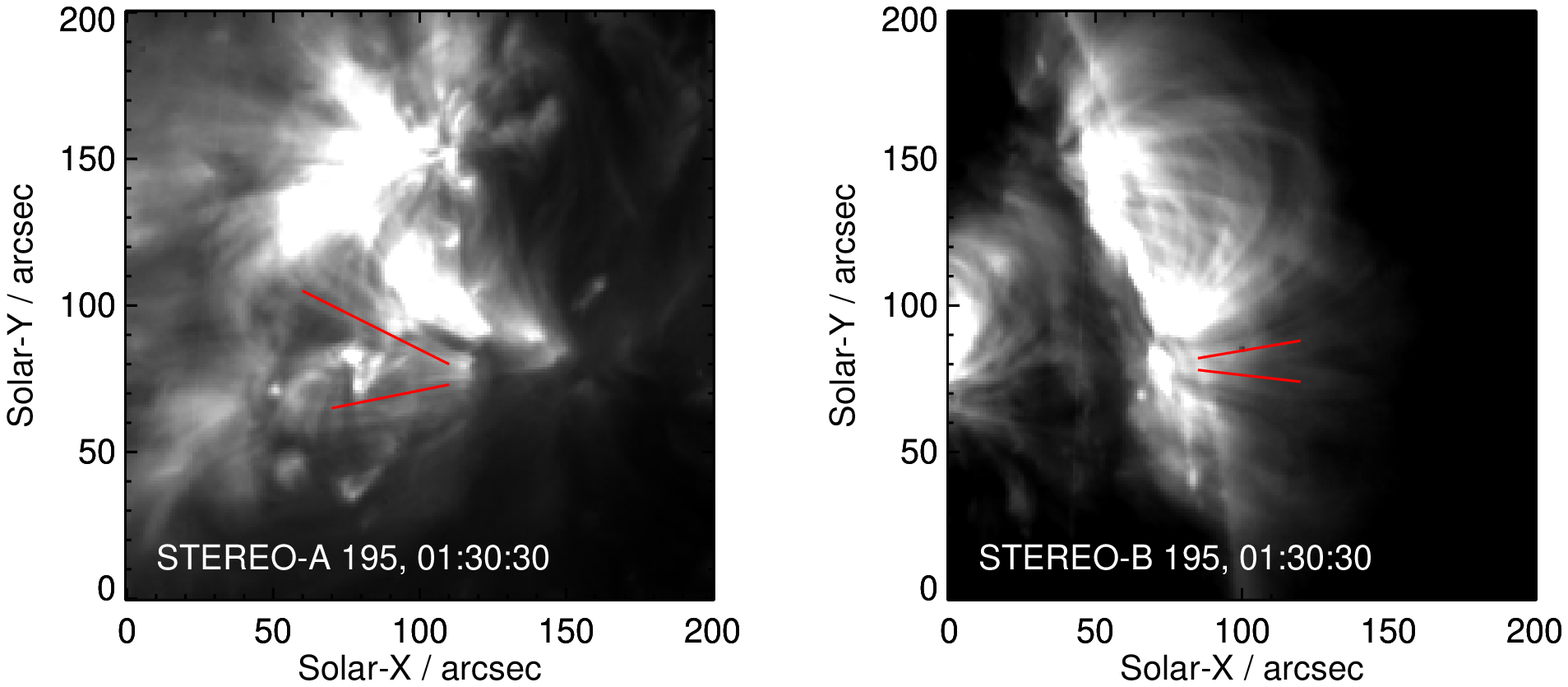}
\caption{STEREO/EUVI 195~\AA\ filter images from the A and B
  spacecraft (left and right panels, respectively) obtained on 2009 November 22
  01:30:30~UT. Red lines indicate the approximate direction of the two
groups of loops that emanate from the footpoint region studied in the
present work.}
\label{fig.st-195}
\end{figure}

Evidence for the different connectivity of loops coming from
footpoints N and S is provided from the STEREO data. The EUVI
instruments of the STEREO A and B spacecraft obtained 195~\AA\ filter
images at a cadence of 2.5 and
5~min, respectively, during the course of the EIS raster. The image frames
from 01:30:30~UT, when EIS was rastering across the footpoint regions,
are shown in Fig.~\ref{fig.st-195}. Compared to the TRACE 171~\AA\ filter
images shown in Fig.~\ref{fig.euvi}, the active region loops
can be seen to extend to greater heights and the footpoint region
considered here is found to connect to two distinct parts of the active
region. The direction of the two groups of loops are indicated by
pairs of red lines overlaid on the Fig.~\ref{fig.st-195} images. The
northerly group of loops connects to a bright part of the active
region, whereas the more southerly loops connect to a fainter part of
the active region. We may speculate that the \ion{Fe}{x} brightness
patch discussed above belongs to the lower group of loops, whereas the
\ion{Fe}{xi}--\ion{Fe}{xii} extended structure belongs to the upper
group of loops.

The above discussion shows that relating \ion{Fe}{xii} emission
structures to \ion{Fe}{viii} structures is very difficult and,
certainly in the present case, not unambiguous. The principal velocity
result from this work relates to \ion{Fe}{viii} and \ion{Fe}{x} for
which the relation between emission structures is very clear
(Fig.~\ref{fig.eis-ims}). However, the velocity derived from
\ion{Fe}{xii} is also of interest due to the presence of an active
region outflow region very close to the \ion{Fe}{viii}
footpoints. This will be discussed in Sect.~\ref{sect.fe12} where
further analysis of the morphology of the \ion{Fe}{xii} structures
will be given.

The following section will present velocities for loop structure B
that are averaged across the loop's diameter, and
Sect.~\ref{sect.dens} will present density and filling factor
estimates for the loop.

\section{Loop structure B}

Fig.~\ref{fig.eis-ims} shows the region used to define the loop
structure B. The spectra along the data columns between (and
including) the two black lines are summed in the Y direction, leading to a single 1D
spectrum for each X-position along the loop. The summation is
necessary both to improve signal-to-noise in weak lines and to correct
any distortions to the line centroid position caused by the
potentially distorted
PSF of EIS (Sect.~\ref{app.psf}).
Spectra at each X-position along the loop were created using the
\emph{Solarsoft} IDL routine EIS\_MASK\_SPECTRUM which averages the
signal from each spatial pixel in the loop cross-section. The emission
lines in the 
resulting 1D spectra were then fit with Gaussian profiles using the
\emph{Solarsoft} routine SPEC\_GAUSS\_EIS to yield intensity, velocity
and line width values for each line.
In total, 43 spectra corresponding to
X-pixels 51 to 95 (pixels 54 and 55 were missing) were fit. In general
we will refer to the X-direction as representing the ``height'' along
the loop, with pixel 51 being the bottom of the loop.

\begin{figure}[h]
\epsscale{0.7}
\plotone{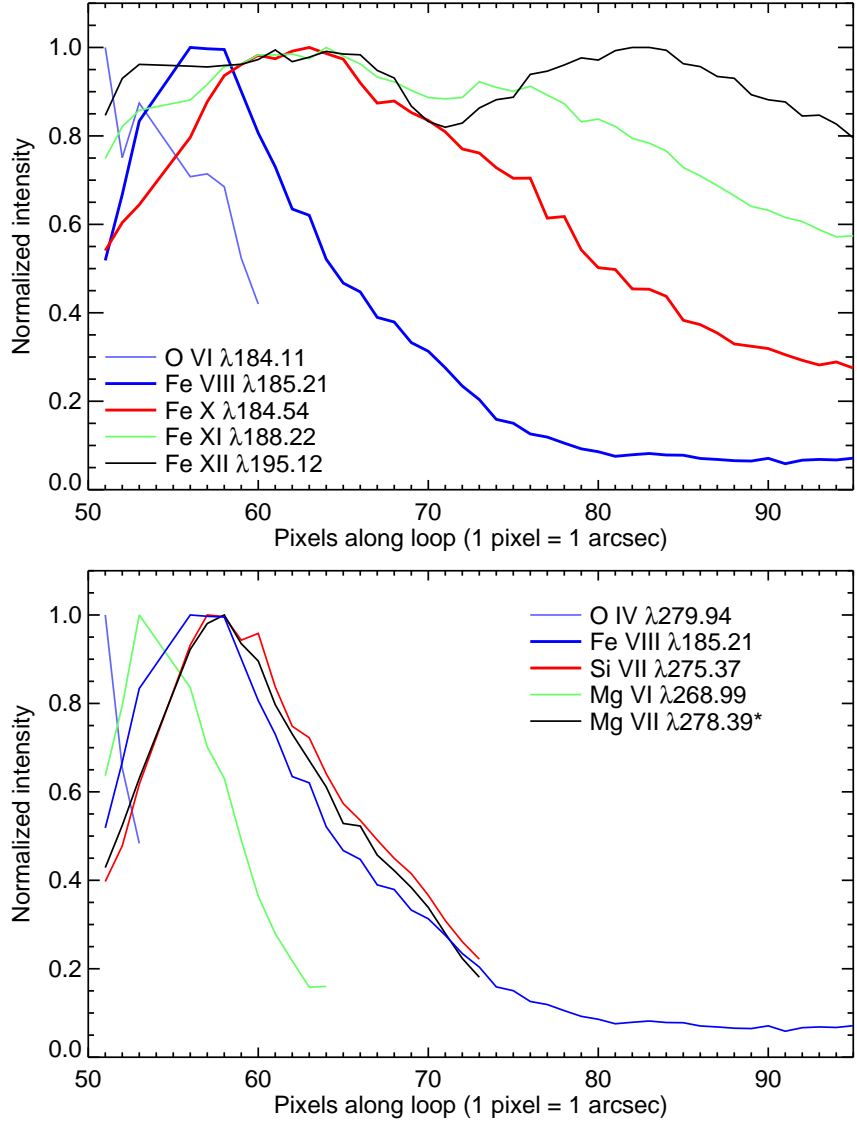}
\caption{A comparison of intensity profiles along loop structure B
  (shown in Fig.~\ref{fig.eis-ims}) for
  selected emission lines from the EIS SW (upper panel) and LW (lower panel)
  channels. The \ion{Fe}{viii} \lam185.21 profile is shown in the
  lower panel for reference. All intensities are normalized to the
  maximum intensity along the loop. The \ion{Mg}{vii} \lam278.39 line
  intensity has not been corrected for a blend with the  weaker
  \ion{Si}{vii} \lam278.44 line.}
\label{fig.loopb-int}
\end{figure}

Fig.~\ref{fig.loopb-int} shows the intensity variation along loop
structure B for nine different emission lines observed by EIS. For the
lines formed at temperatures $\log\,T_{\rm eff}\le 6.05$
(Table~\ref{tbl.ions}) there is a clear correlation between the
height at which the intensity peaks and the temperature. The
two coolest ions, \ion{O}{iv} and \ion{O}{vi}, both peak at the lowest
height and  \ion{Mg}{vi}
peaks two pixels higher up the loop. \ion{Fe}{viii}, \ion{Si}{vii}
and \ion{Mg}{vii} each peak around the same position, three to five
pixels higher than \ion{Mg}{vi}, which is consistent with
their $T_{\rm eff}$ values. The trend continues with \ion{Fe}{x}
\lam184.54 which has a broad peak several pixels higher than \ion{Fe}{viii}.

The intensity profiles for
\ion{Fe}{xi} and \ion{Fe}{xii} are more complex, each showing two peaks
of enhanced emission along the loop. Both peak in intensity at a
location that is a good match to that where 
\ion{Fe}{x} is most enhanced (X-pixels 58 to 66). The second peak for
\ion{Fe}{xi} is around X-pixel 75, and that for \ion{Fe}{xii} is
around X-pixel 83. As discussed in the previous section, our
interpretation is that the two intensity peaks of \ion{Fe}{xi} and
\ion{Fe}{xii} probably represent two groups of loops that head towards
 different parts of the active region, even though they both originate
from footpoint S. The cooler ions (\ion{Fe}{x} and below) are
principally formed
at lower heights where there is less distinction between the two
groups of loops and so a more simple dependence of line intensity with
height is found.

\begin{figure}[h]
\epsscale{0.7}
\plotone{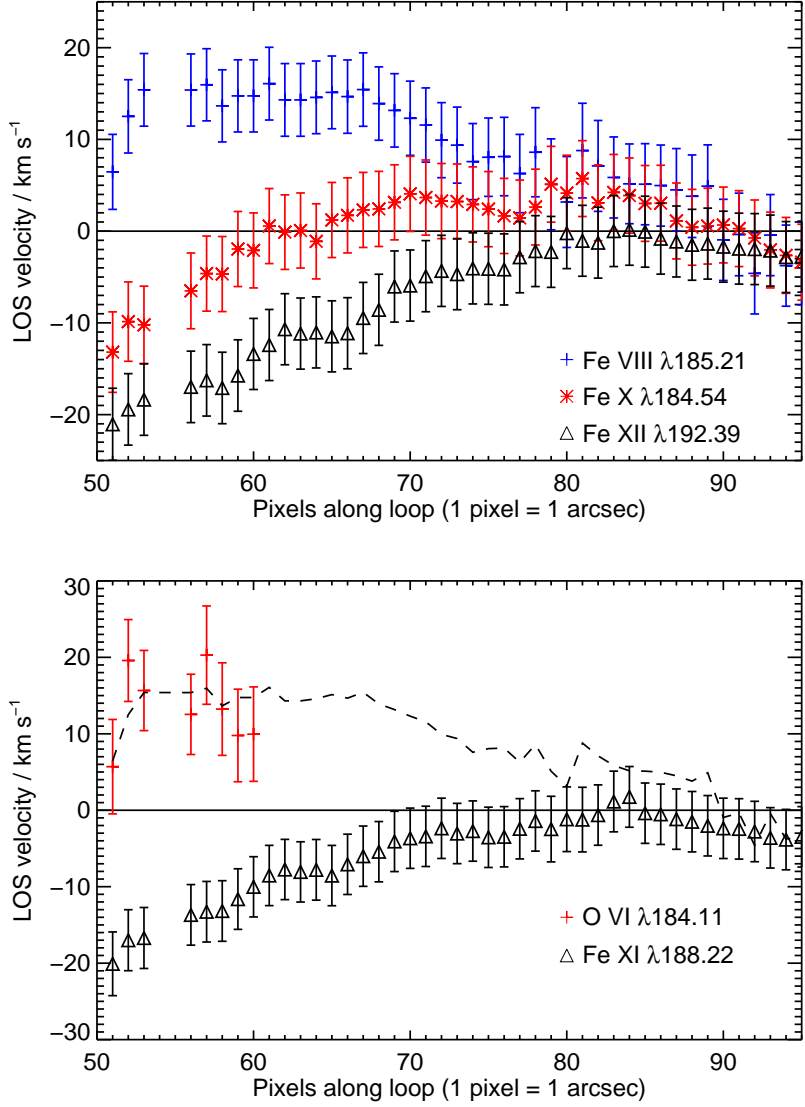}
\caption{A comparison of the velocity profiles of five different
  emission lines along loop structure B (shown in
  Fig.~\ref{fig.eis-ims}). Note the slightly different
  Y-axis scales for each. In the lower panel, the \ion{Fe}{viii}
  \lam185.21 velocities are represented by a dashed line.}
\label{fig.loopb-vel}
\end{figure}

The
velocities along loop structure B are presented in
Fig.~\ref{fig.loopb-vel} which shows the
velocities from  five emission lines within the EIS SW band, displayed over two panels for greater
clarity. The key result is that \ion{Fe}{viii}, over the pixel range
56 to 68, displays a significant redshift of around 15~\kms, yet
\ion{Fe}{x} in the same spatial location is close to rest (there is a
trend of increasing redshift with height from $-5$ to $+3$~\kms).
Taking into
account the error bars, the \ion{Fe}{x} velocity is almost consistent
with being at rest. The cooler \ion{O}{vi} \lam184.11 line can
only be measured accurately in the lowest heights of the loop
structure, but shows values consistent with \ion{Fe}{viii}.
\ion{Fe}{xi} and
\ion{Fe}{xii} show very similar behaviour, with each being close to
rest at X-pixels 70 and above, but showing significant blueshifts in
the lower heights of the loop structure (X-pixels 51--65). \ion{Fe}{x} also displays
significant blueshifts at the lowest heights. Our interpretation of
these blueshifts is that they arise from an  ``outflow
region'' \citep[e.g.,][]{harra08,doschek08,bryans10} for which the
plasma intrudes into the line of sight of loop structure B. This is
discussed in Sect.~\ref{sect.fe12}.

\begin{figure}[h]
\epsscale{0.7}
\plotone{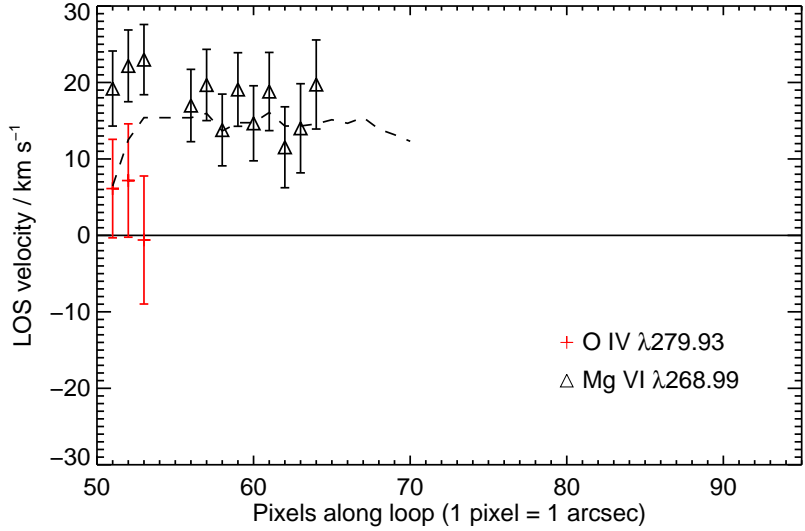}
\caption{A comparison of the velocity profiles of the EIS LW lines
  \ion{O}{iv} \lam279.94 and \ion{Mg}{vi} \lam268.99 along loop
  structure B (shown in Fig.~\ref{fig.eis-ims}).
   The \ion{Fe}{viii}
  \lam185.21 velocities are represented by a dashed line.}
\label{fig.loopb-vel-lw}
\end{figure}

The EIS LW spectrum gives access to additional low temperature lines
and, in particular, access to \ion{O}{iv} \lam279.94 which is the
lowest temperature line observed by EIS that is suitable for velocity
measurements. As discussed earlier in Sect.~\ref{sect.fe8}, the use of
\ion{Fe}{viii} \lam185.21 as a reference line for the wavelength
calibration fails for the LW band, as the separation between the SW
and LW bands varies with time. Using the method described in
Sect.~\ref{sect.lw}, the velocities derived for \ion{Mg}{vi} \lam268.99 and
\ion{O}{iv} \lam279.94 are shown in Fig.~\ref{fig.loopb-vel-lw}. The
\ion{Mg}{vi} line shows good agreement with \ion{Fe}{viii}, except for
the lowest three heights where the \ion{Mg}{vi} velocity is
significantly higher. By contrast, the \ion{O}{iv} line agrees with
\ion{Fe}{viii} at the two lowest heights, but is significantly
discrepant at the third height. Considering the four coolest
ions together, a definitive statement on the velocity profile at the
lowest heights in the loop structure can not be made. \ion{O}{iv},
\ion{O}{vi} and \ion{Fe}{viii} all suggest that the velocity decreases
towards the base, whereas \ion{Mg}{vi} suggests it remains constant,
or slightly increasing. There is, however, no evidence for a
significant increase in velocity towards the lower, cooler part of the
loop footpoint region, i.e., a step increase in velocity occurs between
\ion{Fe}{x} ($\log\,T=6.05$) and \ion{Fe}{viii} ($\log\,T=5.85$), but
there is no further large
velocity increase  between $\log\,T=5.2$
and $\log\,T=5.8$.

\begin{figure}[h]
\epsscale{0.7}
\plotone{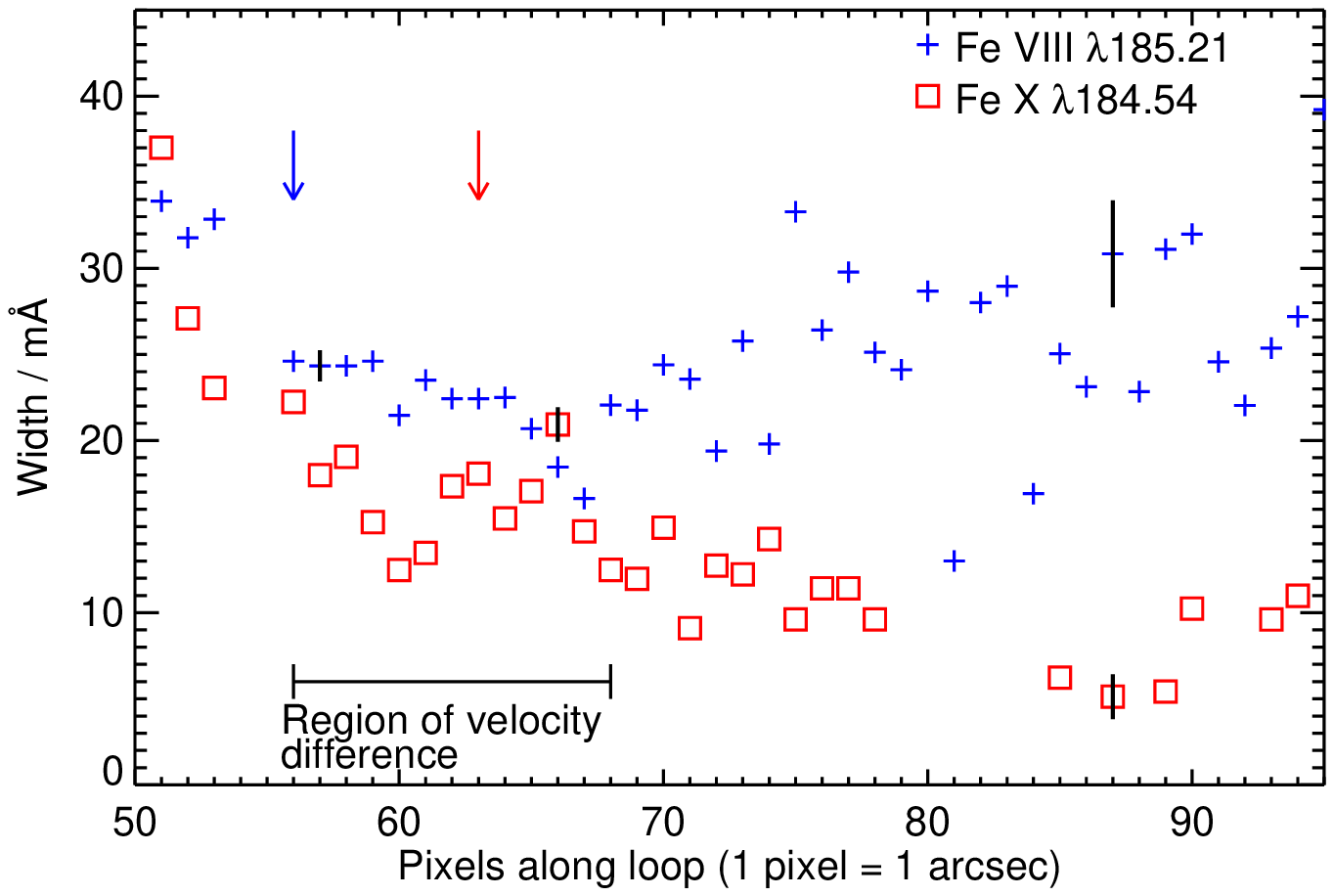}
\caption{Non-thermal line widths of \ion{Fe}{viii} \lam185.21 (blue) and
  \ion{Fe}{x} \lam184.54 (red) along the loop. Error bars on selected
  points are indicated by vertical black lines. The downward-pointing
  arrows indicate the locations of the peak 
  intensities of the two lines. The spatial region where there is a
  significant velocity difference between the two ion lines is
  indicated. Note that some data-points are missing for \ion{Fe}{x}
  \lam184.54 as the measured width was smaller than the combined
  instrumental width and thermal width at these points.}
\label{fig.loopb-wid}
\end{figure}

Returning to the  difference in velocities between
\ion{Fe}{viii} and \ion{Fe}{x}, an important point to note is that it
occurs at the same spatial locations and 
is consistently different over a 13 pixel region. One possible
interpretation is that there could be two plasma components: one that
is redshifted (downflowing) that emits in both \ion{Fe}{viii} and
\ion{Fe}{x}, and one that is hotter with little \ion{Fe}{viii}
emission  that is at rest or blueshifted. In this scenario
\ion{Fe}{viii} is redshifted -- as observed -- while the composite
line profile from the two plasma components in \ion{Fe}{x} would be
significantly less red-shifted, or at rest. This can be ruled out,
however, by considering the emission line
widths along loop B shown in Figure~\ref{fig.loopb-wid}. The
measured full widths at half-maximum have been corrected for the
instrumental width, as given by the \emph{Solarsoft} routine
EIS\_SLIT\_WIDTH \citep[see also][]{young11c}, and the thermal width
for which the $T_{\rm eff}$ values from Table~\ref{tbl.ions} have
been used. The displayed widths therefore represent non-thermal
broadening. Over the X-pixel region 56 to 68, where there is a 15~\kms\
difference between the \ion{Fe}{viii} and \ion{Fe}{x} velocities, the
\ion{Fe}{x} line actually displays a smaller non-thermal broadening
than the \ion{Fe}{viii} line, which is evidence against the
\ion{Fe}{x} emitting plasma having two velocity components\footnote{We
  note that if the $T_{\rm max}$ values from Table~\ref{tbl.ions} are
  used instead of the $T_{\rm eff}$ values, then the non-thermal widths
  of \ion{Fe}{viii} \lam185.21 would increase by about 2--3~m\AA\ over the region
of velocity difference.}. We
therefore believe that there is no significant \ion{Fe}{x} emitting plasma
that is downflowing at speeds of 15~\kms.


\section{Density and filling factor}\label{sect.dens}

The velocity results for \ion{Fe}{viii} and \ion{Fe}{x} imply that
there are multiple, independent components to the loop structure. An
alternative method of studying this is to measure the loop filling
factor with a density diagnostic. \ion{Mg}{vii} is formed at a similar
temperature to \ion{Fe}{viii} and it has an excellent density
diagnostic: \lam280.72/\lam278.39. It has previously been used for
density and filling factor measurements by \citet{young07a},
\citet{tripathi09} and 
\citet{tripathi10}. 

The filling factor at each X-pixel along the loop is determined as
follows. The emission line intensities measured at each location represent the
average intensities over the Y-cross-section. Following the notation
of \citet{tripathi10}, the average intensity, $I_{\rm obs}$, can be converted to a
plasma column depth, $h$, with the expression
\begin{equation}\label{eq.coldepth}
h={ I_{\rm obs} \over 0.83 A(z) G_0(N_{\rm e}) N_{\rm e}^2 }
\end{equation}
where the constant 0.83 is the proton-to-electron ratio in the
corona, $A(z)$ is the abundance of the emitting element relative to
hydrogen, $G_0$ contains a number of atomic parameters \citep[as
described in][]{tripathi10} and is a weakly-varying function of
density, and $N_{\rm e}$ is the electron number density. 

\begin{figure}[h]
\epsscale{1.0}
\plotone{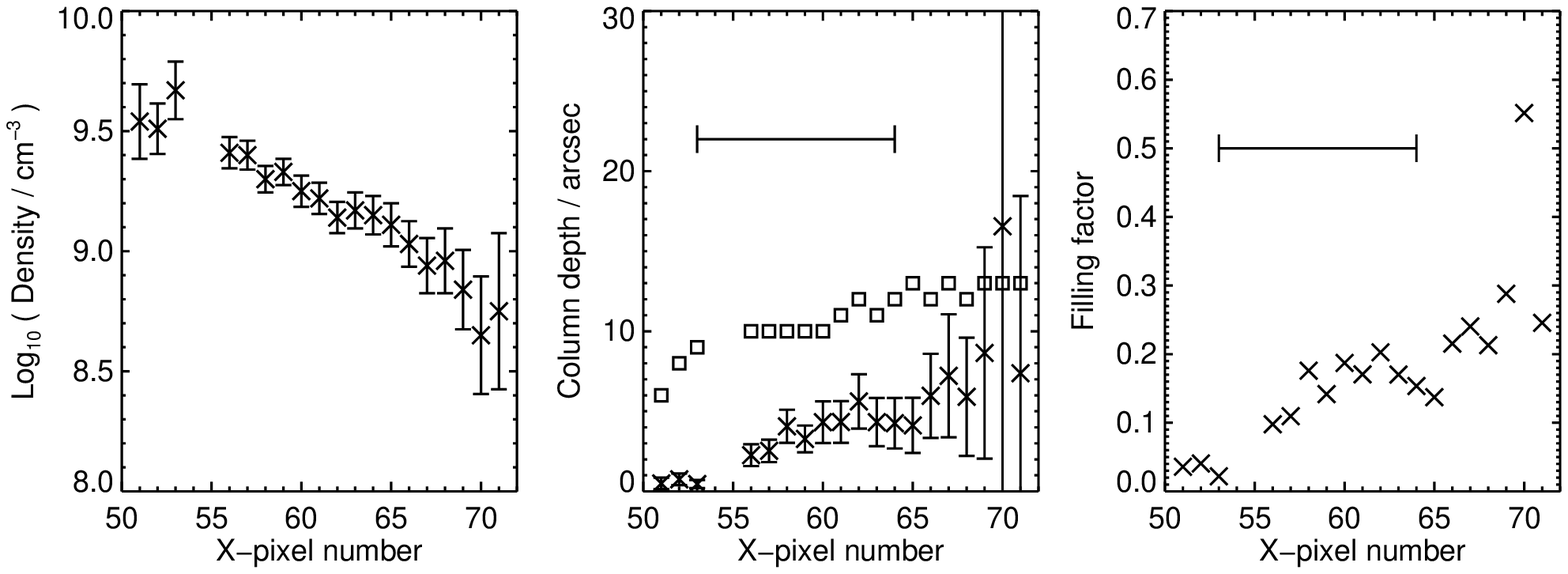}
\caption{Electron densities, column depths and filling factors (left,
  middle and right panels, respectively) as a
  function of position along the loop  derived from the
  \ion{Mg}{vii} \lam\lam278.39, 280.72 line pair. The boxes in the
  middle panel show the observed width of the loop as shown in the
  upper panels of Fig.~\ref{fig.eis-ims}. The pixel region 53--64,
  where \ion{Mg}{vii} is most intense, is
  highlighted in the middle and right panels.}
\label{fig.mg7}
\end{figure}

From the
measured \ion{Mg}{vii} line intensities along the loop, the value of
$N_{\rm e}$ at each X-pixel was determined using atomic data from
the CHIANTI atomic database. A background component to each intensity
was estimated by considering a small region adjacent to the loop, and
this was subtracted from the loop intensities. The derived densities
are shown in Fig.~\ref{fig.mg7} (left panel) where a general trend of decreasing
density with height is seen. Using Eq.~\ref{eq.coldepth} the column
depth at each position can be calculated. For the magnesium abundance
we use the coronal value of \citet{feldman92}, and the atomic parameter
$G_0$ is computed from atomic data in version 6 of the CHIANTI
database \citep{dere09}, including the ionization fractions presented
in that paper. The column depths are presented in the middle panel of
Fig.~\ref{fig.mg7}, where the size of the Y cross-section at each
location is also shown.

To determine a filling factor it is necessary to make some assumption
of the loop geometry. Since the loop is considered to be at an angle
of $\approx20^\circ$ to the line of sight, then the Y-slices at each
X-pixel will be approximately elliptical with the minor diameter,
$d_i$, given
by the width of the loop in the Y direction. Assuming a loop angle of
$\approx20^\circ$ then implies the major diameter (in the
line-of-sight) is $d_i/0.34$. 
The filling factor at X-pixel $i$ is then given by 
\begin{equation}
f_i = { d_i h_i \over \pi (d_i/2)^2/0.34 }  = {1.36 h_i \over \pi d_i}
\end{equation}
The values of $f_i$ are shown in the right panel of Fig.~\ref{fig.mg7}.

In both the middle and right panels of  Fig.\ref{fig.mg7} the pixel
region 53--64 is highlighted as this is where the \ion{Fe}{viii} line
intensity is strongest (Fig.~\ref{fig.loopb-int}). The method used to
determine the column depth essentially requires the emission lines to
be principally formed close to the temperature of maximum
ionization ($T_{\rm max}$). The loop images suggest that the temperature is increasing
with height along the loop, and so the \ion{Mg}{vii} (and
thus \ion{Fe}{viii}) lines will be  brightest where the loop
temperature is close to the ions' $T_{\rm max}$ values. The column depth and
filling factor values should thus be most accurate in this region, and
be less accurate at lower and higher heights where the displayed
values are likely to be underestimates.

Another factor that strongly affects the column depth and filling
factor results is the element abundance. The \citet{feldman92} value
for Mg is a factor four larger than the photospheric value, a standard
enhancement that has been reported from remote sensing observations of
the corona and in situ measurements of the solar wind
\citep{feldman00}. However, fan 
loop structures have been found to display different
enhancements. \citet{young97} found a factor 10 enhancement for a loop
structure that was likely a fan loop, while \citet{widing01} found
that the magnesium enhancement was found to increase with time in fan
loop structures over several days rising to a factor 9 in one case. We
thus note that a larger enhancement of Mg than that of
\citet{feldman92} would reduce the column depth and filling factor
values shown in Fig.~\ref{fig.mg7}.

The \ion{Mg}{vii} filling factor increases with height, with a value
of 0.05 near the base (X-pixels 51--53), and rising to around 0.3
around X-pixel 70. The spatial region where \ion{Fe}{viii} and
\ion{Fe}{x} have different velocities has a filling factor of 
0.1--0.3 at the temperature that \ion{Fe}{viii} is formed. This is 
consistent with the \ion{Fe}{viii} emitting structure only
occupying a fraction of the emitting volume, with the \ion{Fe}{x}
emitting structure occupying some remaining part of the volume.

Loop filling factors from the EIS \ion{Mg}{vii} \lam280.72/\lam278.39 ratio
have previously been presented by \citet{young07a} and
\citet{tripathi09} who both found  filling factors close to 1 in the
regions close to the loops' footpoints. The present loop structure B is known to
have filamentary structure based on the two strands identified from
the TRACE 171 images (Sect.~\ref{sect.loop}), and so smaller filling
factors are not surprising. The loops studied by \citet{young07a} and
\citet{tripathi09} were aligned much closer to the plane of the sky
than the present loop structure which may have aided in identifying a
single monolithic loop.

\section{Interpretation of the emission from Fe\,XII}\label{sect.fe12}

Sect.~\ref{sect.loop} discussed the temperature structure of the
spatial region around the \ion{Fe}{viii} points and highlighted the
complexity of the \ion{Fe}{xii} \lam195.12 image which stands in
contrast to the simple, two footpoint appearance of the \ion{Fe}{viii}
image. In this section, we investigate in more detail the
morphology revealed by \ion{Fe}{xii}.

\begin{figure}[h]
\epsscale{0.7}
\plotone{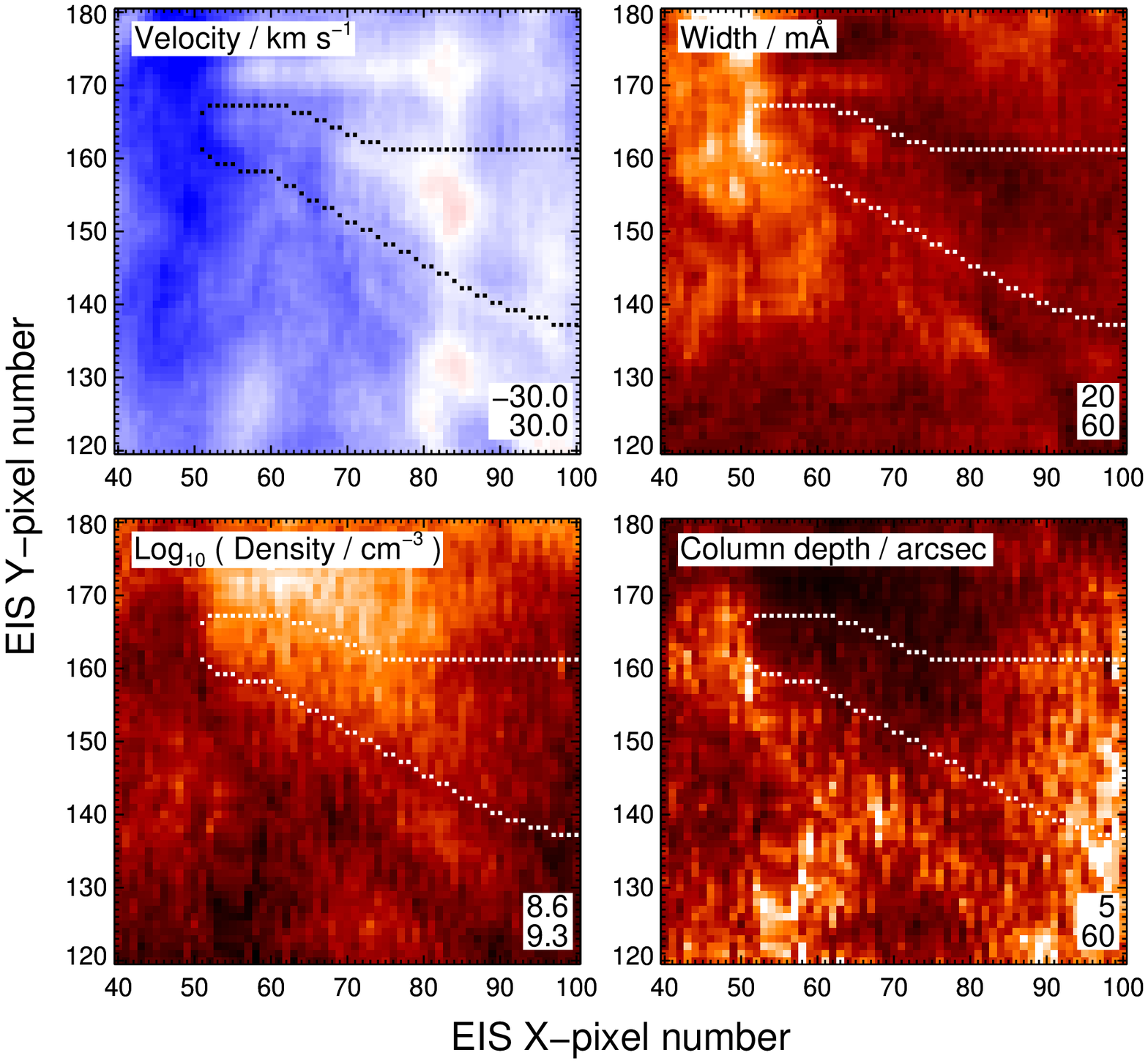}
\caption{Plasma parameter maps derived from emission lines of
  \ion{Fe}{xii} for the same raster sub-region shown in
Fig.~\ref{fig.eis-ims}. The top-left and top-right panels show
velocity and line width maps obtained from the \lam195.12 emission
line. The bottom-left and bottom-right panels show density and column
depth maps derived from the (\lam186.85+\lam186.89)/\lam195.12
ratio. In each plot the loop structure B is indicated by the two
dotted lines. The two  numbers in the bottom-right corner of each panel
give the minimum and maximum value displayed in the images.}
\label{fig.fe12-maps}
\end{figure}

Fig.~\ref{fig.fe12-maps} shows four images derived from the
\ion{Fe}{xii} \lam195.12 line. Each covers the same spatial region as
the \ion{Fe}{xii} \lam195.12 intensity image of Fig.~\ref{fig.eis-ims}. The two upper panels
show velocity and line width maps for \ion{Fe}{xii} \lam195.12, while
the two lower panels show density and column depth maps obtained from
the \ion{Fe}{xii} (\lam186.85+\lam186.89)/\lam195.12 ratio. The
\lam186.85, \lam186.89 lines are blended, forming a single emission 
line in the spectrum. This line and \lam195.12 were fit with the
automatic Gaussian-fitting procedure EIS\_AUTO\_FIT \citep{young11a}
using the prescription of \citet{young09b}. Densities and column depths
were derived from
the measured
\ion{Fe}{xii} ratio using version 6 of the CHIANTI database and the
EIS\_DENSITY procedure \citep{young11b}.

The most striking feature of the \lam195.12 velocity 
map is the blueshifted region to solar-east of the loop footpoints,
which also coincides with larger line widths. Comparing also with the \lam195.12
intensity image of Fig.~\ref{fig.eis-ims} shows that the \lam195.12
intensity is lower in this region. These three features are a
signature of so-called
active region outflow regions
\citep{delzanna08,doschek08,bryans10}. We note that the \lam195.12
profiles from the present region do
not show the distinct emission component on the blue side of the
emission line that was found by \citet{bryans10} for an active region
observed in 2007 December. Instead the line profile is fit well with a
single Gaussian. The overlay of the loop B outline on the
Fig.~\ref{fig.fe12-maps} images shows that the \ion{Fe}{viii} footpoint
actually lies within the \ion{Fe}{xii} outflow region.
Although the blueshifts are strongest at the base of the loop, and
just to solar-east of the base, there are significant blueshifts along
the body of the loop up to X-pixels of around 70, and also the regions
surrounding the loop. This suggests that the outflowing, low intensity
plasma may be intermingled with the brighter loops. Note that at
X-pixels 80--90 the velocity measurements are consistent with no net
line-of-sight flow. We speculate that this region corresponds to loops
and, in particular, the features labeled A--D in
Fig.~\ref{fig.eis-ims}. Along the body of loop B there is a trend of
increasing line width and blueshift towards the footpoint, which may
indicate an increasing contribution to the line of sight emission from
the outflow region plasma.


The density and column depth maps of Fig.~\ref{fig.fe12-maps} reveal
densities of around $\log\,N_{\rm e}$=8.75--9.00 and column depths of
30--60~arcsec for the X-pixel region 80--100. These column depths are  consistent with
the size of the long loop structures that are seen extending from the footpoint region
in the STEREO 195 images of Fig.~\ref{fig.st-195}. Towards the base of
loop structure B (X-pixels 50--75) the densities are in the range
$\log\,N_{\rm e}$=8.95--9.15 and the column depths in the range
12--23\arcsec. Sect.~\ref{sect.loop} suggested that this region
(corresponding to the \ion{Fe}{x} ``brightness patch'') may represent
loops that connect to a different spatial location than the loops in
the X-pixel region 80--100. However, inspection of the STEREO images
of Fig.~\ref{fig.st-195} suggests that both sets of loops have similar
heights and brightnesses which should be reflected in similar density
and column depths. This is not the case and so there may be low-lying
\ion{Fe}{xii} emission that is co-spatial with \ion{Fe}{viii} and
\ion{Fe}{x}, although rather more diffuse than these ions.

In summary, the \ion{Fe}{xii} emission around the loop footpoints is
rather complex. There are large, extended loops with heights of up to
60\arcsec\ that extend away from the footpoints, probably at multiple
angles to the line-of-sight. There is denser, low-lying plasma
surrounding the cool footpoints, and there is outflowing plasma
neighboring, and probably intermingling, with the loops
that is around a factor 2--3 times less intense than the loop emission.

\section{Discussion}\label{sect.discuss}

The most significant result from this work is that a single loop
structure observed with EIS exhibits significantly different line-of-sight
velocities in two emission lines that are relatively close in
temperature. \ion{Fe}{viii} \lam185.21 shows a redshift of $\approx
15$~\kms\, while \ion{Fe}{x} \lam184.54 shows velocities close to 
0~\kms. The differences apply at the same spatial locations in the
loop, and are sustained over a projected distance of 9~Mm along the
length of the loop. An immediate consequence is that the loop consists
of at least two plasma components at different temperatures: one is
cool and downflowing, the other is hot and at rest. The fact that the
velocity difference is maintained over a large spatial distance means
that the two components represent two distinct structures rather than,
e.g., the loop simply showing downflows near the bottom, cooler part
of the loop and stationary plasma in the upper, hotter part of the
loop. The higher spatial resolution of the TRACE instrument revealed
that the EIS loop probably consists of at least two narrow `strands'
visible at 1~MK in the 171 filter
(Fig.~\ref{fig.eis-ims}). However, since both strands are emitting
at 1~MK then they will both contribute to the EIS \ion{Fe}{x}
\lam184.54 emission line and so do not represent the two velocity
components.

This result has important consequences if it
is treated as a general property of fan loops. Consider, for example,
the work of \citet{winebarger02} who found \ion{Ne}{viii} redshifts in
fan loops identified from TRACE 171 images. The authors proceeded to
make a loop model with a steady flow that could reproduce the TRACE
171 intensity profile along the loop. Since \ion{Ne}{viii} is formed
at the same temperature as \ion{Fe}{viii} then the redshift result is
consistent with the present work \citep[although the magnitude of the
velocity was significantly larger in the][loop system]{winebarger02}. The TRACE 171 channel is dominated by \ion{Fe}{x} with a
smaller contribution from \ion{Fe}{ix}. If our EIS result is
extrapolated to the \citet{winebarger02} loop system, then the TRACE
171 emission has \emph{little or no physical relation} to the \ion{Ne}{viii}
emission, and so the \ion{Ne}{viii} velocity result can not be
included in a model that seeks to interpret the intensity variation in
the 171 loop image.

\begin{figure}[h]
\epsscale{0.7}
\plotone{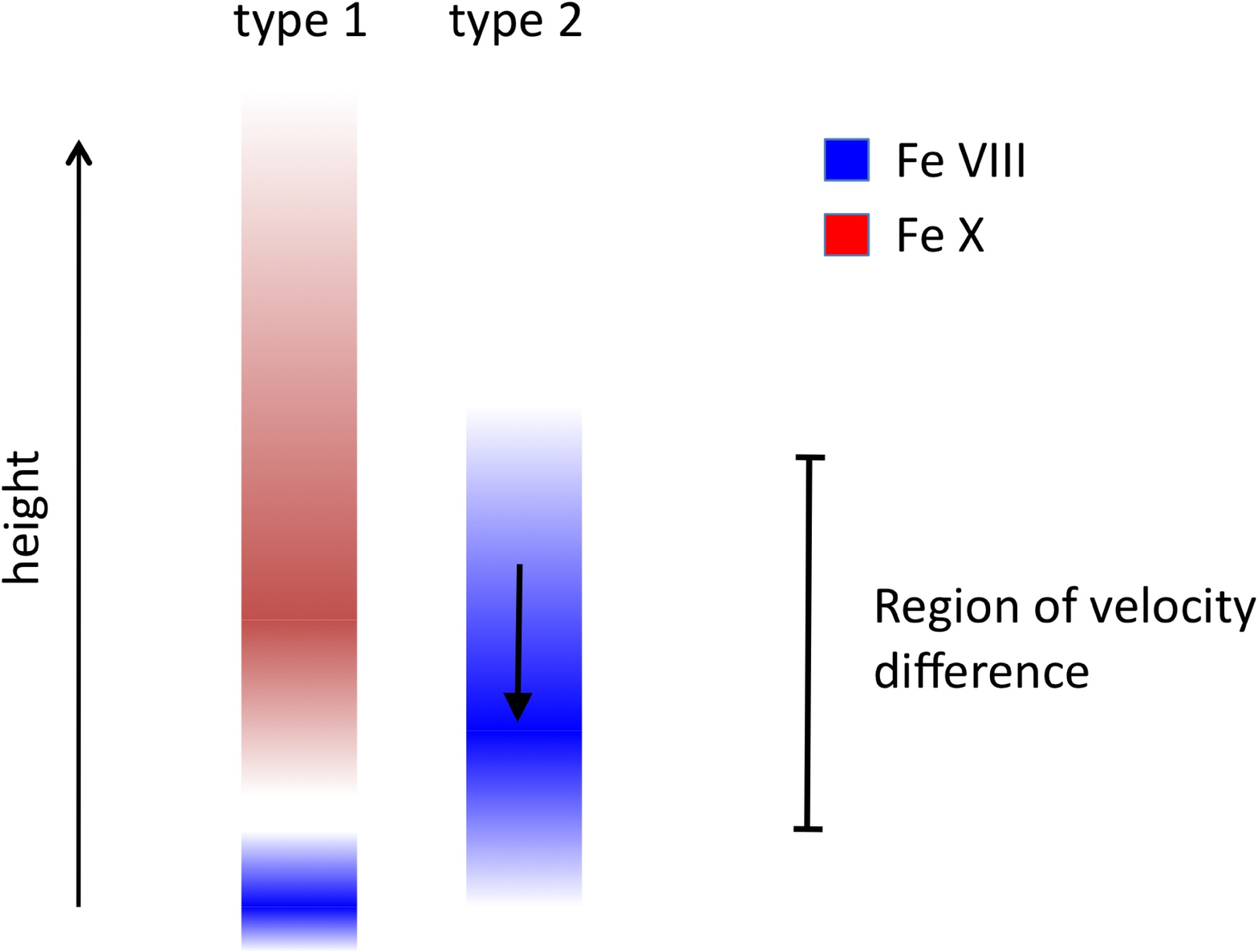}
\caption{A conceptualization of the two types of loop strand that
  could explain the EIS velocity results. See text for more details.}
\label{fig.cartoon}
\end{figure}

One concept for how the loop may be physically structured is shown in
Fig.~\ref{fig.cartoon}. Two distinct loop types (or strands) are
shown. Type 1  is stationary and emits \ion{Fe}{x} along a large portion
of its length, while type 2 is downflowing and emits \ion{Fe}{viii} along a large portion
of its length. The ``centroid'' of the \ion{Fe}{x} emission in the
type 1 strand is located
somewhat higher than that of  \ion{Fe}{viii} in the type 2 strand, as
suggested by the EIS images (Fig.~\ref{fig.eis-ims}). Since
the type 1 strand must cool before descending into the photosphere, it
must also emit \ion{Fe}{viii} emission, but this emission will be much
more compact than that coming from the type 2 strand.
We note that the velocity of \ion{Fe}{viii} \lam185.21 does
fall at the lowest heights in the EIS loop (Fig.~\ref{fig.loopb-vel})
and the width increases (Fig.~\ref{fig.loopb-wid}) which may indicate
that the type 1 strand is contributing significant \ion{Fe}{viii}
emission. 

One question relates to the presence of any emission in type 2 strands
from ions hotter than \ion{Fe}{viii}. It would be surprising if the
strand did not achieve coronal temperatures, yet the coronal emission
lines observed by EIS show that such emission either does not partake
in the \ion{Fe}{viii} flow, or that it is sufficiently weak that it is
masked by the stationary coronal plasma of the type 1 strands. 

The work of \citet{delzanna08}, \citet{tripathi09},
\citet{delzanna09a} and \citet{warren11} suggests that
the velocity results found here may be a common feature of fan loops.
If this is correct then, despite the division of the loop into two types of
independent strand, these two strands always appear together in fan
loops, and their relative distribution/strength always conspires to
make \ion{Fe}{x} appear stationary and \ion{Fe}{viii} downflowing. If
there was significant imbalance towards type 1 strands, then
\ion{Fe}{viii} would no longer display extended emission, instead
being compressed into a small spatial region at the base of the
\ion{Fe}{x} strand. While an imbalance towards type 2 strands would
yield a fan loop with either little \ion{Fe}{x} emission, or with a
\ion{Fe}{x} line profile showing
significant downflow velocity. 
A natural solution to the apparent connectedness of the two strand types
is that the type 2 strand represents a
later, cooling 
stage in the evolution of a type 1 strand. However, the number of
stationary strands is well-balanced with the number of cooling,
downflowing strands in order that a predominance of one over the other
is not seen in the observations.

These arguments are somewhat speculative and require a 
survey of many loops
before they can be confirmed. Certainly a check on
the velocity
difference between \ion{Fe}{viii} and \ion{Fe}{x} needs to be
performed on many loops from different active regions. A thorough
velocity analysis such as the one presented in this work is not
necessary. Simply measuring the wavelength separation of the nearby
\ion{Fe}{viii} \lam185.21 and \ion{Fe}{x} \lam184.54 lines in the
loops and comparing with the standard, off-limb rest separation will
be sufficient. A study in the same loops of the intensity
distributions of the \ion{Fe}{viii} and \ion{Fe}{x} lines with height
would also be profitable by determining the heights of maximum
emission. A time history of how the \ion{Fe}{viii}--\ion{Fe}{x}
velocity difference changes (or does not change) for a loop is also an
important study, as it may reveal if the balance between type 1 and
type 2 strands changes with time.

The concept of apparently monolithic loop structures consisting of
multiple, unresolved strands has been explored previously by theorists,
particularly with regard the nanoflare theory of coronal heating
\citep{parker88,cargill94}, where small heating events only heat
individual strands, but combined can account for global loop
properties such as temperature, density and emission
measure. Multi-stranded loop models have also been cited to explain
observations that appear to show some loops have an isothermal
temperature profile \citep{lenz99,reale00}. 

Dynamic, 1D models of individual loops/strands that include plasma
flows have been made by \citet{spadaro03}, \citet{spadaro06} and
\citet{bradshaw10}. \citet{spadaro03} considered active region loops of 40--80~Mm in
length, with peak temperatures of around 4~MK and coronal
densities of up to $2\times 10^{10}$~cm$^{-3}$; while
\citet{spadaro06} modeled quiet Sun loops with lengths of 5--10~Mm,
peak temperatures up to 1.3~MK, and coronal densities up to
$10^9$~cm$^{-3}$. The parameters of the loop modeled by
\citet{bradshaw10} are closest to  those of the loop studied here -- a loop length of 67~Mm, peak temperature of 3.6~MK and density
of $3\times 10^9$~cm$^{-3}$ -- so we summarize briefly the velocity
results from this work. The loop was allowed to reach a hydrostatic
equilibrium with a coronal apex temperature and cool footpoints, and
then allowed to cool with no heating applied. Enthalpy and conduction
fluxes balance the radiative losses of the plasma, and
\citet{bradshaw10} found that the enthalpy flux dominates in certain
temperature regimes and evolution periods. As the loop progressively
cools, increasing downflows are found near the peak temperature of the
loop, varying from $\approx$10~\kms\ around 2~MK to $\approx$40~\kms\
at 0.3~MK. At the temperatures at which \ion{Fe}{viii} and \ion{Fe}{x}
are formed (0.7--1.1~MK) the velocities are around 15--20~\kms, and
most of the length of the loop shows significant downflows. The
\ion{Fe}{viii} velocity measured in ``loop structure B'' of the
present work is consistent with this model, but the flow is only found
in the lower portion of the loop, and \ion{Fe}{x} shows no significant
velocity. 

The model can not be directly compared with the observation as the
observed loop likely consists of many strands, and so a better
comparison would be with a time-average of cooling strands randomly
distributed in time. Even then, there is likely a contribution to the
emission line intensities and velocities from the heating phase of the
strands which is not modeled by \citet{bradshaw10}.
The \ion{Fe}{viii}--\ion{Fe}{x} velocity result presented here would
seem to be a valuable constraint on loop models of this sort, and
modelers are encouraged to investigate how it can be reproduced.


Finally, a comment on coronal velocity measurements. The present paper
has laid out the difficulties of determining absolute velocities from
the EIS instrument. Although EIS is capable of measuring velocities as
\emph{precise} as 0.5~\kms\ \citep{mariska08}, they are only
\emph{accurate} to, at best, 4~\kms. Improved accuracy at EUV
wavelengths (i.e., below 912~\AA) is unlikely for the forseeable future of solar physics due
to the lack of absolute wavelength standards. The only solution
appears to be observing coronal lines above the Lyman limit where
either a calibration lamp can be used, or photospheric/low
chromosphere lines with small velocities that serve as wavelength
fiducials can be observed. Since the allowed transitions of all coronal ions are found
below 700~\AA\ the emission lines must be observed in second,
third or even fourth spectral order, but then blending can become a
significant issue. There are some coronal forbidden lines that are
found above the Lyman limit but below where the photospheric continuum
begins to become significant ($\approx 1400$~\AA). The most notable
such line is probably 
\ion{Fe}{xii} \lam1242. 


\section{Summary}

A loop footpoint region aligned towards the observer's line-of-sight
has been  identified in ions formed from
$\log\,T=5.2$ to 6.0, showing a clear progression of hotter emission
with greater heights. The continuation of the loop to higher
temperatures of $\log\,T=6.1$ to 6.3 is less distinct and suggests
two groups of loops that may return to the solar surface at different
locations on the opposite side of the active region.
Line-of-sight velocities in one particular footpoint show
redshifts of around 15--20~\kms\ for the ions \ion{O}{vi},
\ion{Mg}{vi} and \ion{Fe}{viii} (formed over $\log\,T=5.5$--5.8). A
clear and distinct drop to approximately zero velocity is found for \ion{Fe}{x}
($\log\,T=6.0$), with the difference in velocity occurring at the same
spatial locations and being maintained over a projected spatial region
of at least 9000~km. \ion{Fe}{xi} and \ion{Fe}{xii}, formed at
$\log\,T=6.1$--6.2, show velocities consistent with \ion{Fe}{x}
although the loop structure is probably contaminated by distinct,
outflowing plasma along the line-of-sight at low heights.

By comparing the line widths of \ion{Fe}{viii} and \ion{Fe}{x} it is
shown that the different velocities of the two ions imply that the
\ion{Fe}{x} emission comes from a different structure (or structures)
than \ion{Fe}{viii} despite the fact that the loops visible in the
ions' images are co-spatial. The solution suggested here is that there
are two distinct types of strand within the loop: one is hot and
stationary, the other is cooler and downflowing. The second strand may
simply represent a later stage of evolution of the first strand, but
in this case there must be many  strands that average out to
give the observed loop emission.

A density diagnostic of \ion{Mg}{vii} was used to demonstrate that the
density falls monotonically with height from $\log\,N_{\rm e}$=9.5 at
the base to $\log\,N_{\rm e}$=8.6 at a projected height of
15,000~km. Assuming a circular cross-section for the loop structure,
the filling factor is estimated to increase from around 0.05 at the
loop base to close to 0.3 at 15,000~km. Over the region where there is a
difference in line-of-sight velocity for \ion{Fe}{viii} and
\ion{Fe}{x} the filling factor is around 0.2. As \ion{Mg}{vii} is
formed at a similar temperature to \ion{Fe}{viii}, this filling factor
result supports the conclusion that the \ion{Fe}{viii} emission comes
from only a fraction of the loop volume, with the \ion{Fe}{x} emitting
strands contributing to some part of the remaining volume.

Inspection of line parameter maps from \ion{Fe}{xii} shows that the
loop structure considered here lies very close to, or even within, an
active region outflow region. These regions appear to be common to
many active regions, and their close proximity to fan loops has
previously been noted by \citet{warren11}. Mixing of the outflow
plasma along of the line-of-sight to the loop structure complicates
velocity measurements for \ion{Fe}{xi} and \ion{Fe}{xii}.

\acknowledgments

The work of P.R.Y.\ was funded by NASA under a contract to the U.S.\ Naval Research
Laboratory.
B.O.D.\ and H.E.M.\ acknowledge STFC (UK). B.O.D. was supported by funding from
the Gates Cambridge Trust. H.~Warren and I.~Ugarte-Urra are thanked
for useful discussions. Hinode is a Japanese mission developed and launched by
ISAS/JAXA, with NAOJ as domestic partner and NASA and
STFC (UK) as international partners. It is operated by
these agencies in co-operation with ESA and NSC (Norway).

{\it Facilities:} \facility{Hinode(EIS)}, \facility{STEREO(EUVI)}, \facility{TRACE}.

\appendix

\section{Comparison between the quiet Sun and HK methods for
  wavelength calibration}\label{app.kamio}

\begin{figure}[h]
\epsscale{0.7}
\plotone{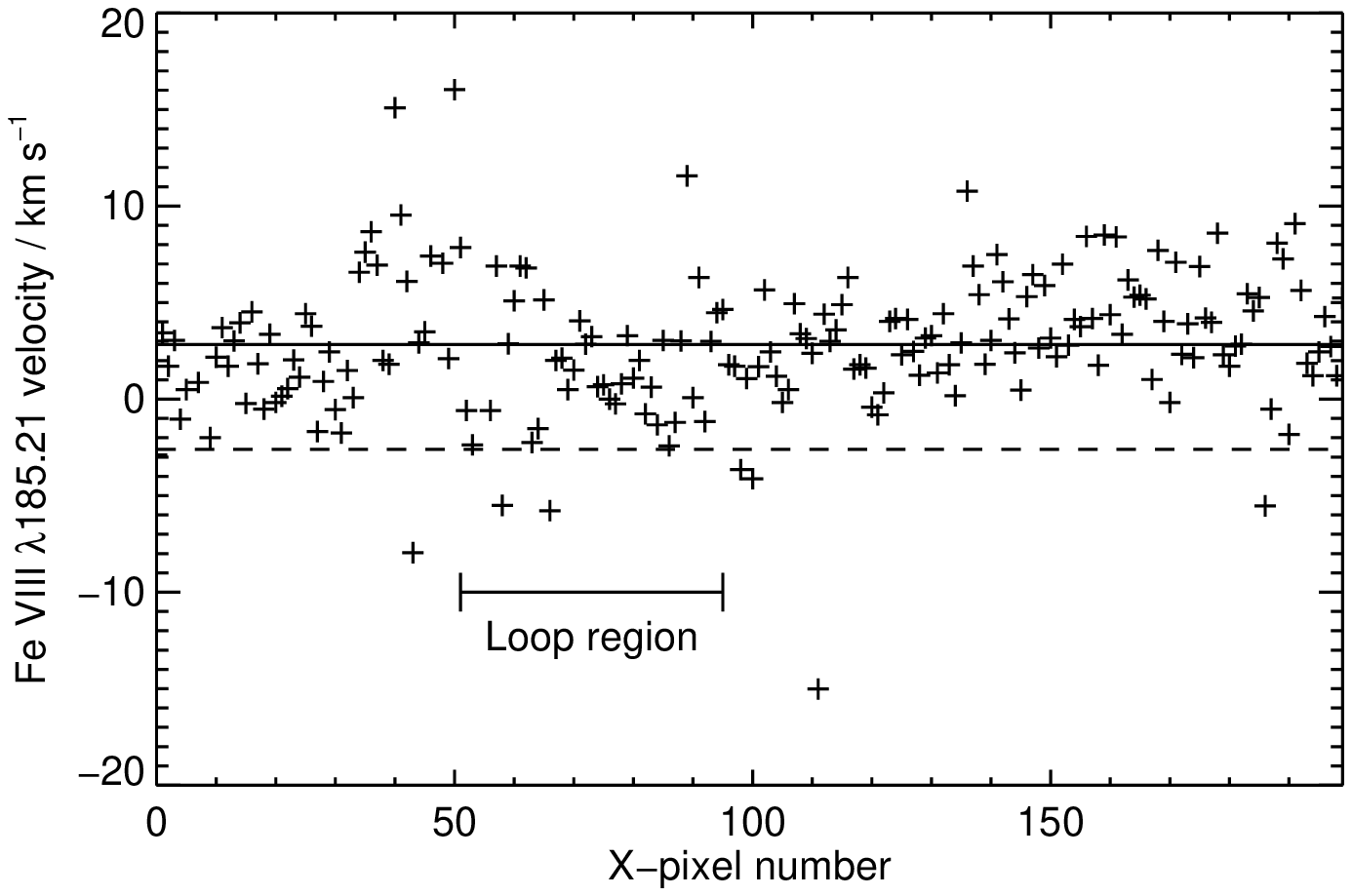}
\caption{Absolute \ion{Fe}{viii} \lam185.21 velocites in the quiet Sun part of
  the 2009 November 11 raster obtained using the method of
  \citet{kamio10}. The dashed horizontal line denotes the quiet Sun
  velocity assumed for the calibration method described in
  Sect.~\ref{sect.fe8}. The X-pixel range that corresponds to the
  coronal loop is indicated.}
\label{fig.kamio}
\end{figure}

The empirical model of \citet{kamio10} allows the absolute wavelength
of the \ion{Fe}{xii} \lam195.12 emission line at an instance in time to be determined purely
from instrument temperature measurements and the motion of
the \hinode\ spacecraft. We can use this method to determine the
average velocity of \ion{Fe}{viii} \lam185.21 in the quiet Sun part of
the 2009 November 21 raster and compare it to the method described in
Sect.~\ref{sect.fe8}. The data are processed in the same way, and so
the bottom 50 pixels of the \lam185.21 raster are averaged to produce
a spectrum with good signal-to-noise. Instead of the orbit variation
being determined from the variation of the \lam185.21 centroid,
however, it is taken from the \citet{kamio10} method. Since this
method corrects the EIS wavelength scale to force \lam195.12 to be at
rest, then it is necessary to apply the off-limb offset of
\citet{warren11} to obtain the rest wavelength of \lam185.21. Upon
doing this, the velocity in the quiet Sun region can be obtained and
the results are shown in Fig.~\ref{fig.kamio}. The
\lam185.21 velocities are mostly redshifts, and the average quiet Sun velocity
over the complete X-pixel range is $+2.8$~\kms, compared to the value
of $-2.6$~\kms\ that was used in Sect.~\ref{sect.fe8}, so there is a
systematic offset of 5.4~\kms\ between the two calibration methods. 
The quiet Sun calibration method used in the present work is preferred
as it is tied to the absolute quiet Sun velocity measurements
of \citet{peter99}, whereas the method of \citet{kamio10} makes the assumption
that the \ion{Fe}{xii} \lam195.12 velocity measured over the life of
the EIS mission averages to zero, which may not be true.

\end{document}